\documentclass[pdflatex,sn-mathphys-num]{sn-jnl}% Math and Physical Sciences Numbered Reference Style 
\usepackage{graphicx}%
\usepackage{multirow}%
\usepackage{amsmath,amssymb,amsfonts}%
\usepackage{amsthm}%
\usepackage{mathrsfs}%
\usepackage[title]{appendix}%
\usepackage{xcolor}%
\usepackage{textcomp}%
\usepackage{manyfoot}%
\usepackage{booktabs}%
\usepackage{amsmath}
\usepackage{algorithm}
\usepackage{algpseudocode}
\usepackage{amssymb}
\usepackage{booktabs}
\usepackage{graphicx}
\usepackage{subcaption}

\usepackage{listings}%
\usepackage[official]{eurosym}
\def\FSR{\boldsymbol{\mathcal{S}_1}}
\newcommand{\FSRi}[1]{\mathcal{S}_{1,#1}}
\def\CSR{\boldsymbol{\mathcal{S}_2}}
\newcommand{\CSRi}[1]{\mathcal{S}_{2,#1}}
\def\Npop{\boldsymbol{N}}
\def\Zcull{\boldsymbol{z}}
\newcommand{\Zculli}[1]{z_{#1}}
\def\NMpop{\boldsymbol{N}_i}
\newcommand{\NMpopi}[1]{N_{i#1}}
\def\ZMcull{\boldsymbol{z}_i}
\newcommand{\ZMculli}[1]{z_{i#1}}
\def\RMulti{\boldsymbol{R}_i}
\newcommand{\RMultii}[1]{R_{i#1}}

\begin{document}

\title[Spatial Joint-Species N-mixture Models for Multi-Source Observational Data with Application to Wild Deer Population Abundance in the Republic of Ireland]{Spatial Joint-Species N-mixture Models for Multi-Source Observational Data with Application to Wild Deer Population Abundance in the Republic of Ireland}


\author*[1]{\fnm{Aoife~K.} \sur{Hurley}}\email{Aoife.Hurley@ul.ie}

\author[2]{\fnm{Ruth~F.} \sur{Carden}}\email{Ruth.Carden@ucd.ie}

\author[3]{\fnm{Sally} \sur{Cook}}%\email{Ruth.Carden@ucd.ie}

\author[4]{\sur{Irish Deer Commission}}%\email{Ruth.Carden@ucd.ie}

\author[5]{\fnm{Ferdia} \sur{Marnell}}%\email{Ruth.Carden@ucd.ie}

\author[6]{\fnm{Pieter~A.~J.} \sur{Brama}}%\email{Ruth.Carden@ucd.ie}

\author[5]{\fnm{Daniel~J.} \sur{Buckley}}%\email{Ruth.Carden@ucd.ie}

\author[1]{\fnm{James} \sur{Sweeney}}\email{James.A.Sweeney@ul.ie}

\affil*[1]{\orgdiv{Department of Mathematics and Statistics}, \orgname{University of Limerick}, \orgaddress{\postcode{V94 T9PX}, \country{Ireland}}}

\affil[2]{\orgdiv{School of Archaeology}, \orgname{University College Dublin}, \orgaddress{\postcode{D04 F6X4}, \country{Ireland}}}

\affil[3]{\orgdiv{School of Geography and Environmental Science}, \orgname{Ulster University}, \orgaddress{\postcode{BT52 1SA}, \country{United Kingdom}}}

\affil[4]{\orgdiv{Irish Deer Commission}, \orgaddress{\country{Ireland}}}

\affil[5]{\orgdiv{National Parks and Wildlife Service (NPWS)}, \orgaddress{\state{Dublin}, \postcode{D07 N7CV}, \country{Ireland}}}

\affil[6]{\orgdiv{School of Veterinary Medicine}, \orgname{University College Dublin}, \orgaddress{\postcode{D04 F6X4}, \country{Ireland}}}


\abstract{
    Accurate estimation of populations and spatial distributions of wild animal species is critical from a management and conservation perspective. It may also be important from an observational study perspective, for example in exploring disease transmission risks between wild and domesticated animals. Where overpopulation of a species is suspected, usually identified by excessive damage to flora or poor herd health, accurate estimates of populations are essential in specifying, justifying, and monitoring the impact of culling programmes. The primary challenge in developing population estimates is typically the sparse and disparate nature of the data sources for model development, collected at differing spatial and/or temporal resolutions. In this article we focus on the challenge of estimating the populations of the three primary deer species in the Republic of Ireland, which are suspected of posing a disease transmission risk to the cattle industry. We present a modelling framework to incorporate sparse point-referenced observational data, aggregated areal-level cull information, and harness expert ecological opinion on likely deer mortality rates  to address the identifiability challenge  between estimating culling and abundance parameters of the proposed model. The spatial distributions of the three species are jointly modelled via a correlated multivariate spatial process with the aim of reducing uncertainty in population estimates by borrowing spatial information across the sparse datasets.  %We conclude the article by providing a framework for the incorporation of citizen science data should it also be available, which provides an approach generalisable to other animal species.
    }




\maketitle

\section{Introduction}\label{sec1}
    For many species of wild animals, accurate estimates of their populations are important for management and conservation, as well as for investigating their impact on the surrounding environment and ecosystems. In the Republic of Ireland there are thousands of fauna species, many of which are mammals~\citep{IrishFauna}, and in recent years the population of wild badgers and wild deer are coming under scrutiny as it is suspected that they may harbour and spread diseases to livestock, in particular bovine tuberculosis~\citep{More:2019ti}. Of further interest in the context of deer is the impact of their browsing and ensuing damage on commercial forestry plantations, which poses challenges to national reforestation targets.  
    
    Species such as deer have no natural predators in the Republic of Ireland, as in many countries, and thus the primary population management strategy available is targeted population culling. Culling programmes are typically used to manage species numbers, to halt the spread of diseases, or to address their negative impacts on habitats.  Localised culling of white-tailed deer, as a disease management strategy has been shown to maintain low disease prevalence~\citep{Manjerovic:2014tr}. Culling programmes are not limited to wild animal species - during the 2001 foot-and-mouth outbreak in the UK, many cattle herds were pre-emptively culled to mitigate disease spread. Tildesley et al.~\cite{Tildesley:2009vk} detail that, while there were aspects of precautionary culling policy that were controversial, the study showed the strategy to be effective with fewer farms losing livestock overall. In many countries, recreational hunters play a role in the management of wild species populations though their numbers may be insufficient to act as a proxy for natural predation, for example recreational hunting alone being insufficient to limit the population growth of wild boar in a number of European countries~\citep{Massei:2015wl}. More generally, it is clear that accurate population estimates are vital to evaluate and potentially influence policies on culling programmes.    
   % Marine mammal culling programmes are often adopted as a fisheries management measure, though often their success has not been evaluated~\cite{Bowen:2013vr}. 

    Among the common challenges in the estimation of wild species populations is the sparse and disparate nature of the data sets typically available for model development. For example, sightings data collected by ecological surveys and other sources such as hunters are typically only reported where animals are observed, introducing issues of bias. Royle~\cite{Royle:2004vg} introduced N-mixture models to model animal populations based on point referenced data and imperfect detection, allowing for the estimation of detection and abundance of a species simultaneously, with repeated observations at sites required to estimate detection probabilities~\citep{Joseph:2009tz, Barker:2022ui}. However, there can be issues of identifiability of parameters in models and computational feasibility when populations are large~\citep{Madsen:2023uv} due to the infinite sums in the N-mixture likelihood framework. 
    Haines~\cite{Haines:2016ul} expresses the N-mixture likelihood in closed form using a hypergeometric function, which is algebraically tractable and computable to a high degree of accuracy obviating the issue of infinite sums in the experienced by Royle~\cite{Royle:2004vg}. Further developments include multi-species extensions of the N-mixture model~\citep{Gomez:2018vw, Mimnagh:2022ux}, to estimate the abundance of rare species and to account for between-species correlations.  
    However, multi-species data have many complexities, including imperfect detection, spatial autocorrelation, between-species correlation, and preferential sampling~\citep{Doser:2023tg}. %Abundance data on wild animal species is collected in numerous ways including the use of motion cameras, tracking devices, pellet identification, capture/re-capture, and site surveys.  Site survey data is generally collected at a fine spatial resolution, such as within a predetermined area or along a predefined path, since collection is limited by cost and human sense (vision and hearing). 

    Another potential complexity includes data being collected at different spatial scales -  spatial misalignment of data adds an additional challenge in combining the data sources for modelling. Statistical downscaling approaches have been used in climate modelling where information is available at a low spatial resolution (from satellites), and is extrapolated to much finer spatial scales.
    There are multiple methods used for statistical downscaling including regression based and neural network based methods~\citep{Benestad:2008up}.
    Pacifici et al.~\cite{Pacifici:2019ty} examined the issue of misaligned data while using integrated species distribution models. Their approach allows predictions to be made at an ecologically relevant scale of inference by leveraging spatial correlation and repeat observations at multiple scales. 
    In disease mapping, spatial disaggregation approaches are used to make fine-scale predictions of disease risk from aggregated response data by using high-resolution covariate data~\citep{Lucas:2021uz}. 
    Simulation studies have shown the performance for various levels of data availability and aggregated area sizes~\citep{Arambepola:2022vt}.
    However, disaggregation involves downscaling areal data to point level data, which does not always align with the available data for a problem.  
    State- and county-level data are incorporated by Hepler et al.~\cite{Hepler:2023wk} when estimating the latent county-level prevalence and counts of people who misuse opioids in Ohio. 
    
    The primary focus of this article is to estimate the populations of three species of wild deer in the Republic of Ireland.
    Other important information such as the spatial distribution of each species and the impact of the differing land covers on each species is also of interest. 
    The paper is structured as follows: we introduce our primary example in Section~\ref{sec:MotivatingExample}. 
    We outline the general model structure for joint species population estimation with spatially misalignment in Section~\ref{sec:PropMod}, where we also comment on parameter inference. 
    Section~\ref{sec:SimStudy} outlines an illustrative simulation study to show the impact of decreasing data availability. 
    In Section~\ref{sec:CaseStudy} we describe the motivating example and case study for this article; the main three wild deer species in the Republic of Ireland.
    We provide the results for both the simulation study and application in Section~\ref{sec:Results}. 
    We illustrate the sensitivity of our population estimates to our choice of cull percentages in Section~\ref{sec:SensitivityToCulls}. 
%    We discuss the addition of data collected from citizen science projects in Section~\ref{sec:CitizenScience}, and how this would filter into the modelling framework proposed.
    Finally, in Section~\ref{sec:Discussion} we discuss our findings and possible avenues for future work. 

%%%%%%%%%%%%%%%%%%%%%%%%%%%%%%%%%%%%%%%%%%%%%%%%%%%%%%%%%%%%%%%%%%%
%%%%%%%%%%%%%%%%%%%%%%%%%%%%%%%%%%%
        
\section{Motivating Case Study: Wild Deer Populations in the Republic of Ireland} \label{sec:MotivatingExample}
    In this article we consider data collected on three wild deer species in the Republic of Ireland: fallow (\textit{Dama dama}), red (\textit{Cervus elaphus}), and sika (\textit{Cervus nippon}). Fallow and sika deer were introduced to the Republic of Ireland in the 1200's and in 1860 respectively~\citep{CARDEN:2011vm, Beglane:2018vj}.
    The population of red deer in County Kerry are descended from a 5,000 year old introduction, whereas other red deer populations in Ireland are descended from introductions from UK and elsewhere in the 19th century to modern times~\citep{CARDEN:2011vm, Carden:2012wo}.
    The wild deer population in the Republic of Ireland has recently been the focus of substantial coverage in the national media due to the proposed link in disease transmission between deer and livestock~\citep{IrishTimesMay2015, Kelly:2021wz} and the potential for collisions on rural roads~\citep{Liu:2018uf}.
    Their negative impact on various habitats including conservation ecosystems, agricultural lands, commercial forestry plantations and semi/native woodlands has also been identified~\citep{WoordlandsofIre, CARDEN:2011vm, Murphy:2013ur}. 
    More recently, the record number of culls from the 2021/2022 season based on hunter returns has led to the highlighted that ``the overall size of the deer population in Ireland is unknown because no census of numbers has ever been conducted"~\citep{IrishTImes:Jan23, IrishTimes:May23}. 

    Disease transmission between wild deer and livestock in the Republic of Ireland is an area of growing concern due to the speculated link to the transmission of bovine tuberculosis~\citep{More:2019ti, Kelly:2021wz}. 
    Previous studies have examined deer as possible transmitters of diseases in the UK~\citep{Gibbs:1975tz, Bohm:2007th}. 
    A recent literature review was conducted on diseases impacting both deer and livestock in Australia~\citep{Cripps:2019ud}.
    More recent modelling strategies into the understanding of the spread of bovine tuberculosis in Ireland contains proxies for deer~\citep{Madden:2021wq}, and an emphasis has been placed on understanding the spread of bovine tuberculosis from deer-to-cattle contact~\citep{More:2019ti, Griffin:2023uk}. 

    Wild deer can cause a variety of damage, through bark stripping, browsing of lateral and leader shoots, and adult males thrashing trees with their antlers. 
    Such damage may affect the individual tree’s overall growth, health (pathways for diseases and pests), and timber quality and yields. 
    Currently there are limited studies in Ireland assessing the negative impact, both economic and on biodiversity, on plantation forestry and both semi-native and native woodlands.
    These studies have however determined estimations of costs of deer damage to Sitka spruce plantations, other commercial plantations, and deciduous woodlands (both commercial and (semi-)native) annually to the Irish economy. 
    At a national level, these costs are estimated to be in the range of \EUR{}1.3 million~\citep{Murphy:2013ur}. 
    However, data was only available from a few locations across the country and there is an absence of quantitative baseline data, thus this annual cost estimate may be lower or higher. 

    Multiple studies have modelled the distribution and range expansion~\citep{CARDEN:2011vm, Murphy:2023uh} and the presence and relative abundance~\citep{Morera-Pujol:2023wj} of wild deer in Ireland.
    Other studies have investigated the degree of hybridisation between sika and red deer~\citep{McDevitt:2009un, Smith:2014tm}, the origins~\citep{McDevitt:2009un, Carden:2012wo, Beglane:2018vj}, and the overlap between wild deer species and endangered or vulnerable plant species~\citep{OMahony:2023vs}.
    One previous study estimated the population of one species of wild deer, sika, using hunter returns~\cite{Kelly:2021wz}. 
    However, in this study they fail to account for the uncertainty in the estimated cull percentages.
    
    At this present time we are not aware of any modelling approach that coherently aggregates the available data resources with respect to their attendant sources of uncertainty to estimate the explicit populations of these three species in the Republic of Ireland.  To the best of our knowledge, no such comparable model exists in any other country either.
    Accurate estimates of the populations of the three deer species are vital for investigating forest health, conservation of habitats, and disease transfer between wildlife and livestock, with the potential to inform policies on culling programmes.  The main goals of this analysis are to (a) estimate, with quantifiable uncertainty, the population of the three deer species in Ireland, (b) assess the spatial distribution of each species and (c) investigate the impact of land types on each species' presence and abundance.

%%%%%%%%%%%%%%%%%%%%%%%%%%%%%%%%%%%%%%%%%%%%%%%%%%%%%%%%%%%%%%%%%%%%

\section{Modelling Framework}\label{sec:PropMod}
    In this section we present a modelling framework for species population estimation in the presence of spatially misaligned data. We first introduce our modelling framework in the single species case, as this provides the foundation for our multi-species model.  %We explain the general components of our model in the case of the single species model Section~\ref{subsec:SingleSpecies}, with specifics regarding the joint multi-species model discussed in Section~\ref{subsec:MultiSpecGen}. 
    
    \subsection{Single Species Spatial N-mixture Model} \label{subsec:SingleSpecies}
       Typically the data sources available on species are spatially misaligned consisting of both point-referenced/small area data and much coarser regional level data. We begin by modelling the point-referenced data first, subsequently incorporating information available at coarser areal levels. 
        
        \subsubsection{High Resolution Spatial Data}\label{subsec:GridMod}
        
            Consider there are $m$ distinct spatial locations, $\FSR = \{ \FSRi{1}, \FSRi{2}, \dots, \FSRi{m} \}^\top$, where predictions of species populations are desired. While observations may be gathered at a point-referenced level, typically the data is presented at an arbitrarily fine-scaled aggregated spatial resolution, say on 1km $\times$ 1km grids or similar.
            In the following, we denote the unknown populations of the species at each location by $\Npop =\{ N_1, N_2, \dots, N_m \}^\top$ corresponding to each location in $\FSR$ - here we ignore the potential of species population moving and being double counted. %The species is observed $o$ times across $\FSR$, and these observations or sightings are denoted by $\boldsymbol{y} = \{ y_1, y_2, \dots, y_o \}^\top$. 
            
            %Repeated observations at individual $\FSRi{i}$ are required to estimate detection probabilities when using the N-mixture model framework. 
            We follow the convention in Madsen and Royle~\cite{Madsen:2023uv}, in assigning a binomial distribution  for number of population members observed in a given instance. We assign a negative binomial distribution for the unobserved total species population at a given location as it allows for various N-mixture models to be applied, as previously shown by Goldstein et al.~\cite{Goldstein:2022va}. Sites may be visited multiple times - let $n_{jt}$ represent the reported species numbers at location $j$ at time $t$. We make the simplifying assumption that species populations do not change across repeated visits to a site - this may be appropriate where the hunting season reflects a small window of opportunity and we assume that population members culled are replaced by newer members from births. Our model at this spatial scale is then represented by:
            
            % \begin{equation} \label{eq:NMixGrid}
            %     \begin{split}
            %         y_{k} &\sim \mathrm{Binomial}(N_{k \in i}, p_{k \in i}, [\theta_w]) \\
            %         N_{i} &\sim \mathrm{Negative Binomial}(\lambda_{i}, [\theta_b]) \\
            %         \text{logit}(p_{i}) &= \delta_{0} + G_{i}^\top\boldsymbol{\delta} \\
            %         \log(\lambda_{i}) &= \beta_{0} + X_{i}^\top\boldsymbol{\beta} + \phi_{i}\text{.}
            %     \end{split}
            % \end{equation}

            \begin{equation} \label{eq:NMixGrid}
                \begin{split}
                    n_{jt} &\sim \mathrm{Binomial}(N_j, p_j) \\
                    N_{j} &\sim \mathrm{Negative \,\,Binomial}(\lambda_{j}, \theta) \\
                    \text{logit}(p_{j}) &= \delta_{0} + G_{j}^\top\boldsymbol{\delta} \\
                    \log(\lambda_{j}) &= \beta_{0} + X_{j}^\top\boldsymbol{\beta} + \phi_{j}\text{.}
                \end{split}
            \end{equation}
 
            The detection probabilities for area $j$ are represented by $p_j$, and the mean abundance of the species in area $j$ is given by $\lambda_j$. Here $\boldsymbol{X}$ and $\boldsymbol{G}$ represent environmental covariates available at each site. 
            Lastly, we include our spatial component through $\boldsymbol{\phi}$. 
            For point-referenced information models for continuous spatial variation such as Gaussian process priors on the spatial structure may be used. For aggregated or areal data, conditional autoregressive (CAR) priors such as the intrinsic CAR (ICAR)~\citep{Besag:1974vs}, or the proper CAR (PCAR)~\citep{Stern:2000vt} can be specified.
            An alternative to CAR priors are simultaneously autoregressive model (SAR), which differs from a CAR prior in its covariance structure. 
            The imposed spatial correlations from both the CAR and SAR models are discussed in Wall~\cite{Wall:2004ux}.

        \subsubsection{Coarser Areal Scale} \label{subsec:RegionalMod}
            We consider the case where indirect information on species populations may also be available at coarser spatial resolutions, for example through regional culling programmes or via hunter reported kills at regional levels for licence renewal. This introduces spatial misalignment with the observational data recorded at two different spatial resolutions. 
            This coarser scale with $r$ units, $r << m$, is expressed using $\CSR = \{ \CSRi{1}, \CSRi{2}, \dots, \CSRi{r} \}^\top$. 
            At this spatial level, indirect population data are observed once at each of the $r$ locations. 
            This data is represented by $\Zcull = \{ \Zculli{1}, \Zculli{2}, \dots, \Zculli{r} \}^\top$.
            Using elements seen in Section~\ref{subsec:GridMod}, our model at this lower resolution spatial level is given by
            \begin{equation} \label{eq:RegionalMod}
                \begin{split}
                    R_{k} &= \sum_{j \in k} \Npop_{j} \\
                    z_{k} &\sim \mathrm{Poisson}(R_{k}\kappa_{k})\text{,}
                \end{split}
            \end{equation}
            where $R_k$ is the population of the species in region $k$ of $\CSR$. The reported data, $z_k$, is assumed to reflect the proportion $\kappa_k$ of the $R_k$ animals in $\CSRi{k}$ that have been culled. While the numbers culled reflect a fraction of a sum constrained population, we propose a Poisson approximation to the binomial likelihood for computational stability and efficiency of sampling based inference strategies, reflecting our experience in fitting such models. Where the proportion of animals culled is a small fraction of the population we can expect this approximation to work well.

            In the proposed framework, we combine equation~\ref{eq:NMixGrid} and equation~\ref{eq:RegionalMod}. 
            This leads to an identifiability issue if both the detection probabilities, $\boldsymbol{p}$ and the regional detection probabilities $\boldsymbol{\kappa}$ are to be learnt from the data. 
            That is, we cannot estimate both abundance and culling rates simultaneously. 
            However, if prior information is known about either the detection probabilities or regional detection probabilities, such as possible intervals, these can be sampled and used as an input to the framework, resulting in the learning of the other. 
            
            % This topic is further discussed in Section~\ref{subsec:CullPercentagesUncert}.  
            % \textcolor{red}{discuss the identifiability challenge of not being able to estimate abundance and culling rates simultaneously}

    \subsection{Extension to Multiple Species}\label{subsec:MultiSpecGen}
        %Having outlined our model for the single species case in Section~\ref{subsec:SingleSpecies}, we will now extend this framework to the multi-species case. 
        We retain the notation from the previous section, Section~\ref{subsec:SingleSpecies}, but with the addition of the subscript $i$, $i$ indicating the species being modelled.
        %This subscript indicates the species being modelled: here we restrict the number of species to three . 
        Therefore, the populations for species $i$ at each of the $m$ locations is given by $\NMpop = \{ \NMpopi{1}, \NMpopi{2}, \dots, \NMpopi{m} \}$. 
        %Each species can have differing numbers of observations, that is $o_{a1} \neq o_{a2} \neq o_{a3}$, and as such the observed data are expressed as $\boldsymbol{y}_a = \{ y_1, y_2, \dots, y_{o_a} \}^\top$. 
        We then rewrite Equation~\ref{eq:NMixGrid} as
        % \begin{equation}\label{eq:MultiFineScale}
        %     \begin{split}
        %             y_{a, k} &\sim D_{w}(N_{a, k \in i}, p_{a, k \in i}, [\theta_{a, w}]) \\
        %             N_{a, i} &\sim D_{b}(\lambda_{a, i}, [\theta_{a, b}]) \\
        %             \text{logit}(p_{a, i}) &= \delta_{a, 0} + G_{i}^\top\boldsymbol{\delta}_a \\
        %             \log(\lambda_{a, i}) &= \beta_{a, 0} + X_{i}^\top\boldsymbol{\beta}_a + \phi_{a, i}\text{.}
        %     \end{split}
        % \end{equation}
                \begin{equation}\label{eq:MultiFineScale}
            \begin{split}
                    n_{ij} &\sim \mathrm{Binomial}(N_{ij}, p_{ij}) \\
                    N_{ij} &\sim \mathrm{Negative \,\,Binomial}(\lambda_{ij},\theta_i) \\
                    \text{logit}(p_{ij}) &= \delta_{0i} + G_{j}^\top\boldsymbol{\delta}_i \\
                    \log(\lambda_{ij}) &= \beta_{0i} + X_{j}^\top\boldsymbol{\beta}_i + \Phi_{i, j}\text{.}
            \end{split}
        \end{equation}
        Environmental covariates $\boldsymbol{X}$ and $\boldsymbol{G}$ are not species dependent, however the parameterisaton allows the coefficients to vary across species.
        In our multi-species model, we include between-species correlation in the formulation of the multivariate spatial surface $\boldsymbol{\Phi}$. 
        In the case of using a multivariate ICAR, or MICAR, $\boldsymbol{\Phi}$ takes the form
        \begin{equation} \label{eq:phIMulti}
            \begin{split}
                \boldsymbol{\Phi} &\sim \text{MVN}(0, Q^{-1}) \\
                Q &= \Sigma \otimes (D - A) \text{,}
            \end{split}
        \end{equation}
        where $\otimes$ represents the Kronecker product, $A$ is the adjacency matrix of $\FSR$, $D$ is the corresponding diagonal matrix where each element is the number of neighbours, and $\Sigma$ is the $m \times m$ positive definite matrix and can be interpreted as the non-spatial precision (inverse dispersion) matrix between species. 
        In the case of a multivariate ICAR, or MICAR, the diagonal elements of $\Sigma$ would be the species specific spatial precision $\boldsymbol{\tau} = \{\tau_1^2, \tau_2^2, \dots, \tau_i^2\}$, and the off diagonal elements would contain a parameter $\rho$ to address the additional spatial correlation between species. 
        In the case of using a multivariate PCAR, or MPCAR, Equation~\ref{eq:phIMulti} would be updated to include the additional parameter $\alpha$, which may vary by species. 
        Details of the MPCAR can be found in Jin et al.~\cite{Jin:2007uu} and Banerjee et al.~\cite{Monograph_HMASD}.

        With the inclusion of our coarser spatial data in $\CSR$, the regional population for species $i$ in region $k$ is given by $\RMulti = \{ \RMultii{1}, \RMultii{2}, \dots, \RMultii{r} \}^\top$ with indirect observed data $\ZMcull = \{ \ZMculli{1}, \ZMculli{2}, \dots, \ZMculli{r} \}^\top$. 
        Adapting Equation~\ref{eq:RegionalMod} results in 
        \begin{equation}\label{eq:MultiRegionalLevel}
            \begin{split}
                R_{ik} &= \sum_{j \in k} \NMpopi{jk} \\
                z_{ik} &\sim \mathrm{Poisson}(R_{ik}\kappa_{ik})
            \end{split}
        \end{equation}
        We allow for the regional detection probabilities $\boldsymbol{\kappa}_{ik} = \{  \kappa_{1k},  \ldots,  \kappa_{ik} \}$ to vary between species in a region $k$, that is the cull proportion within a region is not held constant, $\kappa_{1k} \neq  \kappa_{2k} \neq \ldots \neq \kappa_{ik} $. 

    \subsection{Model Fitting and Inference} \label{subsec:Inference}
        We adopt a Bayesian paradigm, where model specification is complete after assigning a prior distribution $p(\boldsymbol{\Theta})$ for the parameter vector $\boldsymbol{\Theta}$, which represents all unknown model parameters. 
        All models outlined in this article were fitted using the statistical software \verb+R+~\citep{RSoftware}, in particular using the \verb+nimble+ package~\citep{Valpine:2017vg} which has many benefits for ecological applications.
        \verb+nimble+ relies on Markov Chain Monte Carlo (MCMC) algorithms to obtain samples from the posterior distribution, with a similar syntax to both \verb+BUGS+~\citep{Spiegelhalter:2007th} and \verb+JAGS+~\citep{Plummer:2003wk}, making it easy to use, and allows for the customisation of sampling methods.
        Ponisio et al.~\cite{Ponisio:2020va} showcase the customisable nature of \verb+nimble+ for a suite of models, while Lawson~\cite{Lawson:2020tl} provides a tutorial on how to use \verb+nimble+ for disease mapping, including various spatial models for both univariate and multivariate cases.
        
        In our experience, \verb|RStan| is limited in its capabilities for N-mixture models~\citep{RStan}. 
        This is due to issues surrounding the binomial $N$ parameter being an integer-constrained random variable, which is a function of an underlying Poisson count generating process.  
        As noted by Madsen and Royle~\cite{Madsen:2023uv}, the Integrated Nested Laplace Approximation approach (\verb|INLA|~\citep{Rue:2009tw}) is also limited in its capabilities for N-mixture models due to the requirement of averaging survey-level covariates to the site or site-sampling level. 
        Furthermore, \verb|INLA| is not a compatible inference approach for applications that involve data collected at an aggregated areal level, which is present for our case study of wild Irish deer.
        \verb|JAGS|~\citep{Plummer:2003wk} is an alternative inference approach; however, we encountered issues with parameter mixing in the joint 3 species model, mainly regarding issues in specifying Wishart priors for the spatial surface $\boldsymbol{\Phi}$. 

        As previously noted, we cannot estimate both the detection probabilities $\boldsymbol{p}$ and regional detection probabilities $\boldsymbol{\kappa}$. 
        For our motivating example, we can gain more information regarding the cull percentages $\boldsymbol{\kappa}$ compared to the detection probabilities $\boldsymbol{p}$, so therefore we sample $\boldsymbol{\kappa}$ and learn $\boldsymbol{p}$. 
        To determine realistic intervals for the cull percentages, we incorporate expert opinions and 5-year trends in both culled numbers and number of deer culled per licence. 
        We pool the posterior samples from the sampled cull percentages, although this does not exactly correspond to the samples drawn from the posterior distribution when sampling both $\boldsymbol{\kappa}$ and $\boldsymbol{p}$, it is intuitively similar to marginalising out the cull percentages $\boldsymbol{\kappa}$. 
        
        % \textcolor{red}{Perhaps mention weak information in the available data and "borrow" reviewer comments about this being a monte carlo approx} \textcolor{magenta}{@James, I'm not clear as to what additional info is being looked for. } \textcolor{blue}{Reviewer1 comment 7 and Reviewer2 comment 5 is what I'm referring to in terms of what they are looking for }

%%%%%%%%%%%%%%%%%%%%%%%%%%%%%%%%%%%
\section{Simulation Study} \label{sec:SimStudy}

    As a proof-of-concept, a simulation study is used to assess the impact of declining data retention on the performance of the proposed modelling framework. 
    To mimic our motivating example, we simulated data for three species at $m = 625$ sites, with $r=5$ aggregated areas. 
    That is, $\CSR = \{ \CSRi{1}, \CSRi{2}, \dots, \CSRi{5} \}^\top$ and $\FSR = \{ \FSRi{1}, \FSRi{2}, \dots, \FSRi{625} \}^\top$. 
    We simulated a covariance matrix that mimics the compositional nature of the covariates used in our motivating example. 
    Across the three species and $r=5$ aggregated areas within each simulated data set, we kept the regional detection probabilities $\kappa = 0.2$. 
    The spatial dependence parameter $\alpha$ was the same for all 5 generated data sets ($\alpha = 1$), we did however vary the spatial dependence parameters $\boldsymbol{\tau}$ to be reflective of the differing ranges of the three species in our motivating example.

    We simulated 5 data sets, with each generated data set having 8 levels of observation retention. 
    The original simulated data sets each had 3 observations at each of the $m$ sites, that is 1,875 observations per species for our base level 100\% data retention. 
    We explored retaining 50\%, 40\%, 30\%, 20\%, 10\%, 5\%, and 2.5\% of the base level data set to investigate the impacts on population and parameter estimates.
    As N-mixture models require spatially replicated counts to estimate the detection probabilities, at the lower levels of data retention we ensured there were repeated visits.

% \begin{algorithm}[H]
%     \caption{Simulation Strategy}
%     \label{alg:AMH}
%     \begin{algorithmic}[1]
%         \State \textbf{Simulation Parameters:} Choose $x_0$, $m_0$, $c_0$, and $\sigma_d$, $\beta$
%         %\State \textbf{Output:} Unbiased sample from $\pi(x)$ for $n = 1,2,...,N$
%         \For{n = 1, 2, ..., N} 
%             \State Sample $x^{*}$ $\sim$ $\mathcal{N}$($x_n$,$\sigma_n$ $c_n$)
%             \State Apply MH criteria $\alpha$($x^*$,$x_n$) = min [1, $\frac{\pi(x^*) q(x_n | x^*)}{\pi(x_n) q(x^*| x_n)}$]
%             % \If{$\alpha$($x^*$,$x_n$)}
%             %     \State $x_{n+1}$ = $x^*$
%             % \Else
%             %     \State $x_{n+1}$ = $x_n$
%             % \EndIf
%             \State Update the covariance matrix 
%             \State $c_{n+1}$ = $c_n$ + $\frac{1}{n+1}$ (($x_{n+1}$ - $m_n$)$(x_{n+1} - m_n)^{T}$ - $c_n$ ) 
%             \State Update the mean vector 
%             \State $m_{n+1}$ = $\frac{n}{n+1}$ $m_n$ + $\frac{1}{n+1}$ ($x_{n+1}$ - $m_n$)
%         \EndFor
%     \end{algorithmic}
% \end{algorithm}

    For each data set in each scenario, we ran three chains each of 40,000 samples, with a burn-in of 20,000 samples and a thinning rate of 20, resulting in a total of 3,000 MCMC samples. %for each of the six alternative models.
    We fit all models using the \verb|nimble| R package~\cite{Valpine:2017vg}. 
    We assess the performance by comparing the 95\% credible interval coverage rates the overall population for each species, and examining the root mean square error (RMSE) for the estimated spatial surface. 
    
    % Set of simulations to show what happens as the amount of data available reduces.
    \subsection{Results} \label{subsec:DeerSimResults}
        %In this section we examine the impact of reducing the number of observations available on model estimates. 
        % Five data sets with three observations per site were generated and then these data sets were sampled such that 50\%, 40\%, 30\%, 20\%, 10\%, 5\%, and 2. 5\% of the observations were retained. 
        This simulation study was conducted using a DELL XPS 15 9570 laptop with 6 2.9 GHz Dual-Core Intel i9-8950 HK processors and 32 GB of memory, with each level of data retention taking approximately 55 minutes. 
        Thus, approximately 440 minutes of computation time was required for each data set. 

\begin{figure}[h]
            \centering
            \includegraphics[width = \textwidth]{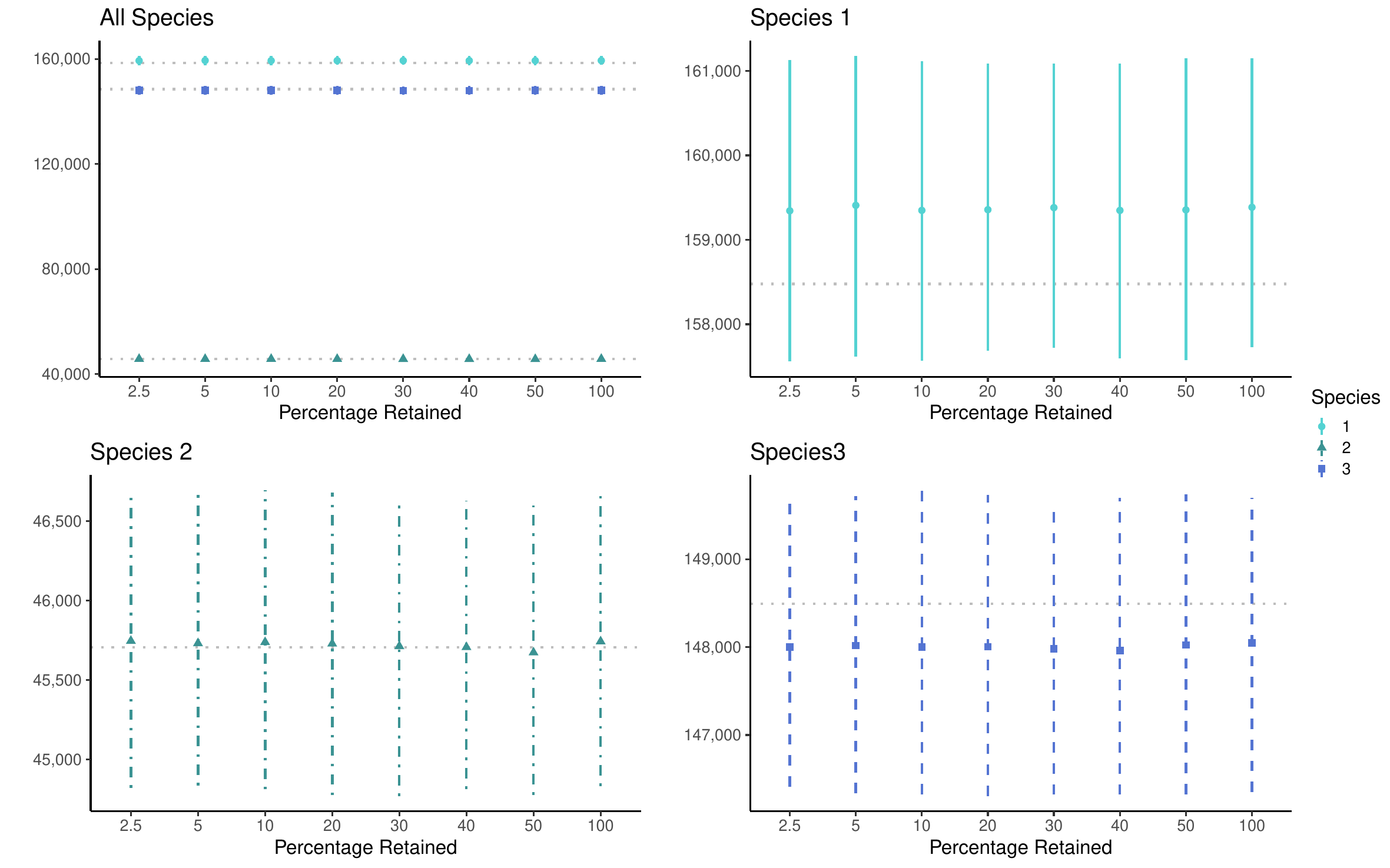}
            \caption{Estimates of three populations with reducing data for one simulated data set. The grey dotted lines indicate the true underlying populations.}
            \label{fig:NTotal_DS1}
        \end{figure}

\begin{figure}[h]
            \centering
            \includegraphics[width = \textwidth]{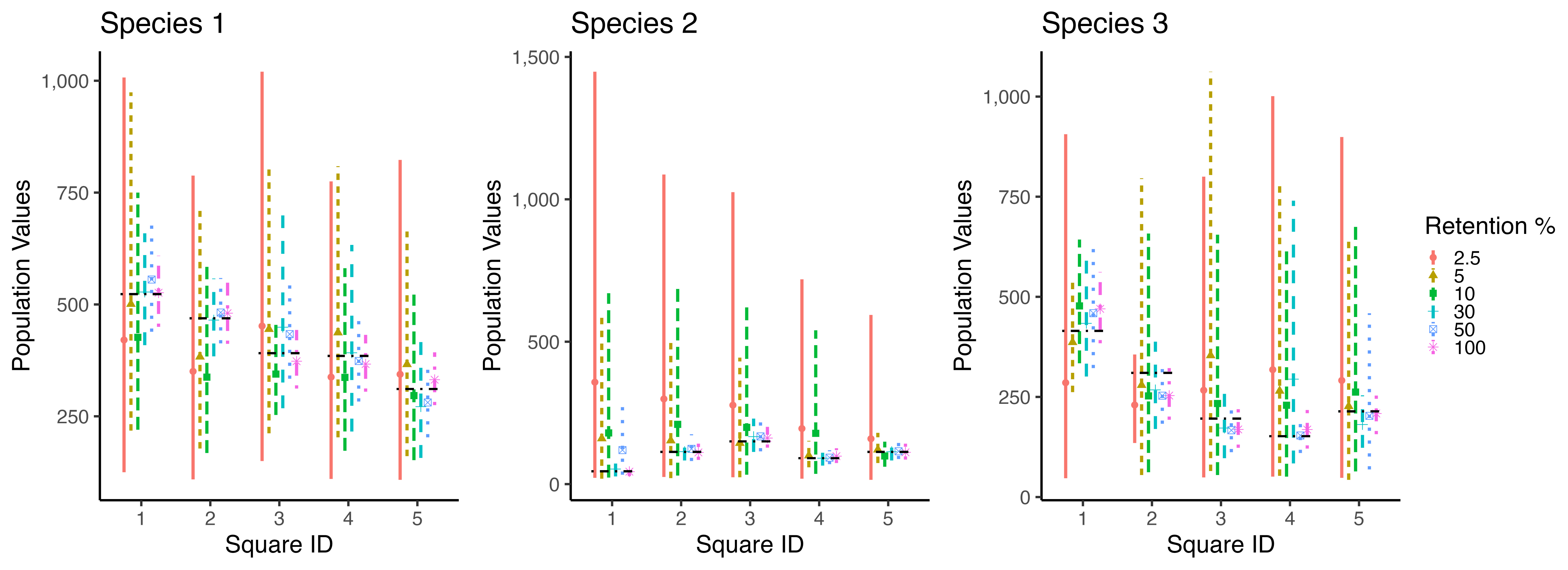}
            \caption{Estimates for the three simulated species at the first 5 grid squares for 6 of the 8 data retention levels. The black horizontal line indicates the true underlying population in the square. }
            \label{fig:NSquare_1}
        \end{figure}

        We are first and foremost interested in gaining accurate estimates of the overall populations of each species. 
        % In these studies, we kept the regional detection probabilities, or cull percentages $\boldsymbol{\kappa}$, constant. 
        By keeping the regional detection probabilities constant, this limits the additional uncertainty we expect to see in these estimates. 
        Figure~\ref{fig:NTotal_DS1} illustrates these estimates and the associated 95\% credible interval for one simulated data set. 
        From Figure~\ref{fig:NTotal_DS1} there appears to be little variation in the estimated population numbers for each species. One reason for the lack of variation is that the regional detection probabilities, $\boldsymbol{\kappa}$, are kept constant at the true underlying value. We examine the impact of sampling $\boldsymbol{\kappa}$ in Section~\ref{subsubsec:VaryCullSims}. 
        Figure~\ref{fig:NTotal_DS1} only shows the overall estimated populations, we do however see greater variability for the estimated species populations at the finest spatial scale. 
        For visual purposes, Figure~\ref{fig:NSquare_1} only shows the estimated populations for the three species in the first 5 grid squares for 6 data retention levels. 
        While Figure~\ref{fig:NTotal_DS1} shows little variation at the overall level, there is quite clearly an increase in variation with a decrease in the percentage of observations retained at the more granular level as seen in Figure~\ref{fig:NSquare_1}. 
        
\begin{figure}[h]
            \centering
            \includegraphics[width = \textwidth]{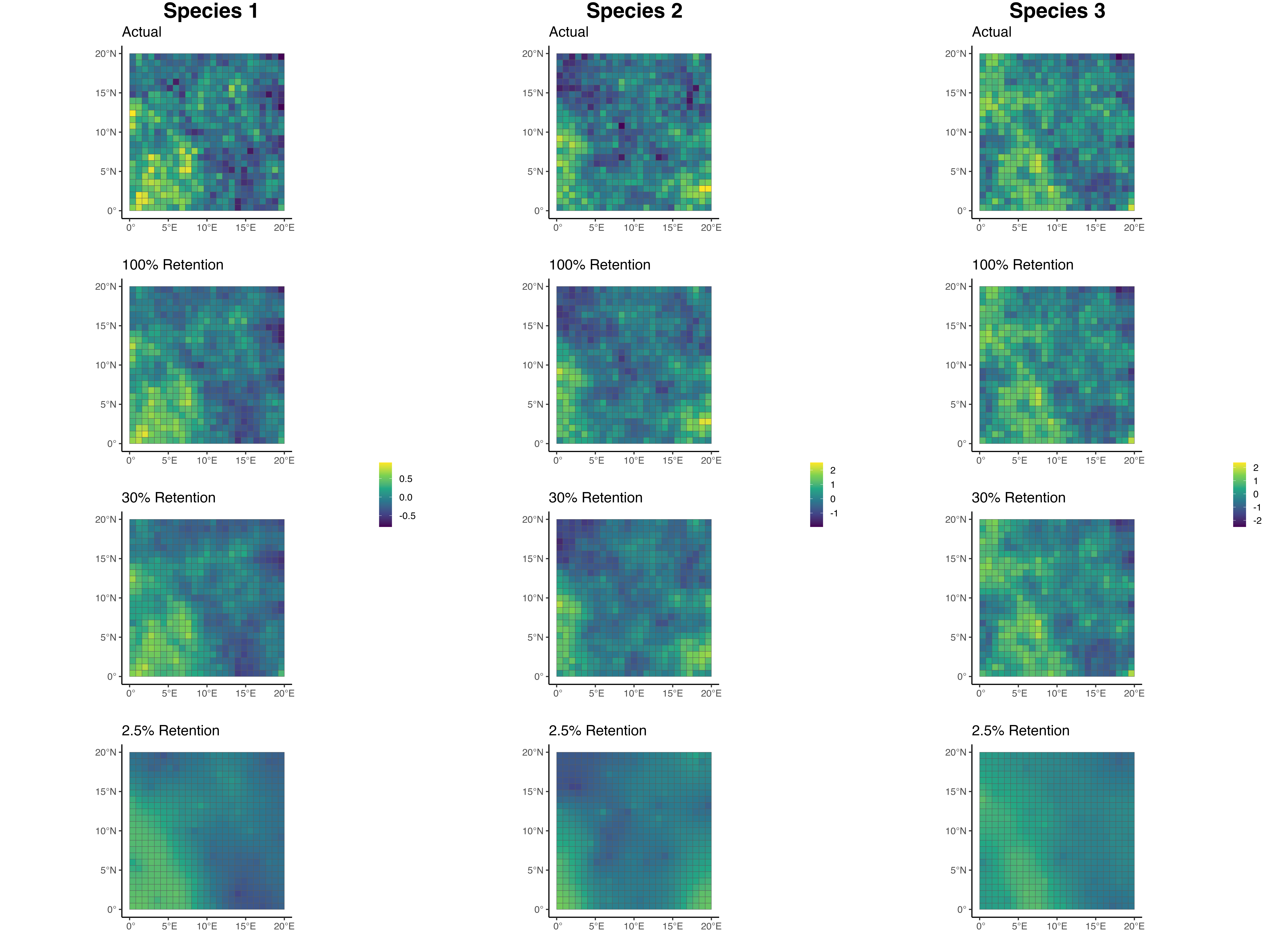}
            \caption{Example of mean surface estimate with three different retention values. As the level of data retention reduces, the mean estimates of the spatial surface are oversmoothed.}
            \label{fig:SurfEstSims}
        \end{figure}

        We are also interested in estimating the underlying spatial surface for each species. 
        Therefore, we examine the RMSE for the underlying spatial surface and visually inspect the estimated surfaces. 
        The overall pattern for the individual spatial surface of each species was captured in the mean estimates, with an example shown in Figure~\ref{fig:SurfEstSims}, although oversmoothing is visible with lower levels of data retention.
        Table~\ref{tab:RMSE1} shows the RMSE for each species for the retention levels shown in Figure~\ref{fig:SurfEstSims} for one data set. 
        The complete set of tables for each data set can be seen in Appendix~\ref{sec:A1}. 
        \begin{table}[h]
            \caption{The RMSE for three retention levels for one simulated data set.}\label{tab:RMSE1}
            \centering
            \begin{tabular}{@{}l c c c@{}}
                \toprule
                Retention Level & Species 1 & Species 2 & Species 3 \\
                \midrule
                100\%     & 0.117  & 0.235 & 0.182 \\
                30\%      & 0.160  & 0.343 & 0.315 \\
                2.5\%     & 0.226  & 0.517 & 0.590 \\
                \botrule
            \end{tabular}
        \end{table}

\begin{figure}[h]
            \centering
            \includegraphics[width = \textwidth]{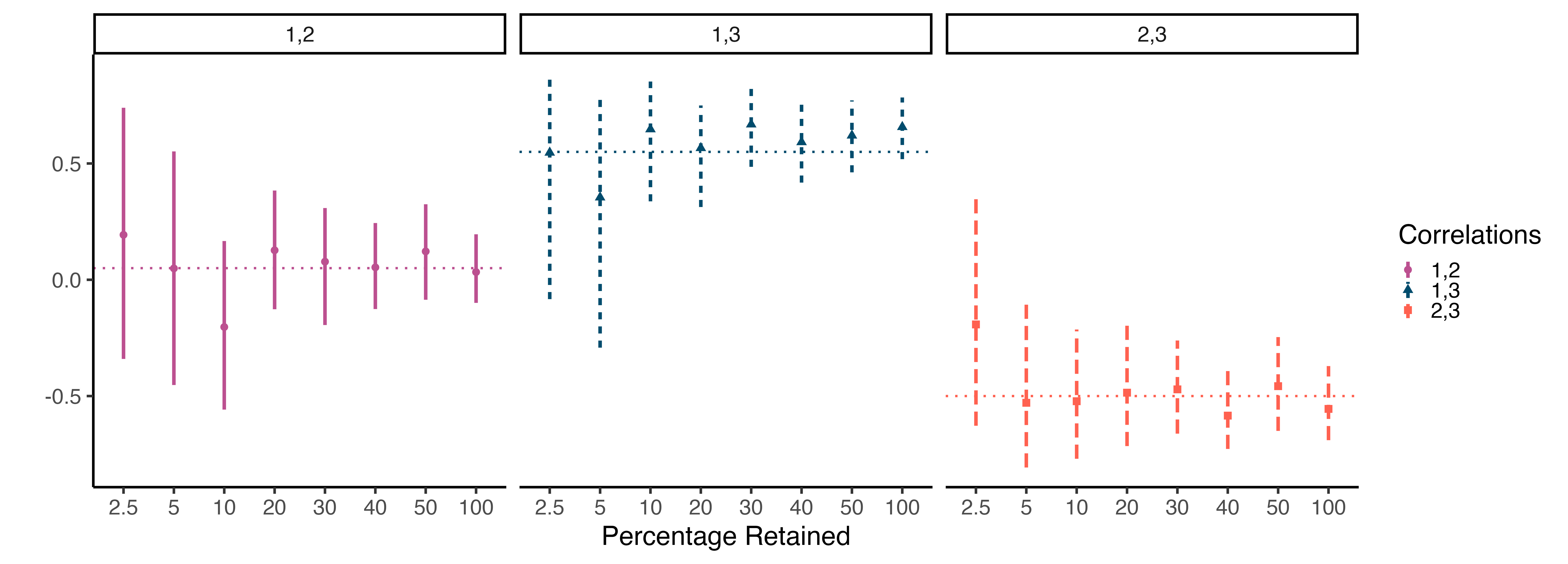}
            \caption{Example of estimated between-species correlations with a reduction in data retention levels. Dotted horizontal lines represent the true between-species correlations. }
            \label{fig:CorrsSims}
        \end{figure}

        As our species are speculated to be correlated, we investigate the impact of reducing data availability on the estimates produced. 
        The underlying simulated values are reflective of what deer ecology experts have witnessed during their observational studies. 
        Similar to the last plots, we visualise this impact for one of the five generated data sets in Figure~\ref{fig:CorrsSims}.
        As the percentage of data retains reduces, the 95\% credible intervals get wider and at the lowest level of data retention all three include 0. 
        
        Lastly, we examine the coverage provided by the 95\% credible intervals. 
        We gain good coverage, even at low data retention level. 
        However, at these lower data retention levels, the variability in many of the parameters is quite large. 
        The 95\% credible interval coverage for all data sets and eight retention levels are shown in Table~\ref{tab:95Levels}. 
        % \begin{table}[h]
        %     \caption{The 95\% credible interval coverage for all five simulated data sets and all data retention levels.}\label{tab:95Levels}
        %     \centering
        %     \begin{tabular}{@{}l c c c c c@{}}
        %         \toprule
        %         Retention Level & Simulation 1 & Simulation 2 & Simulation 3 & Simulation 4 & Simulation 5 \\
        %         \midrule
        %         100\%   & 96.76\%  & 96.52\% & 95.41\% & 95.11\% & 95.96\% \\
        %         50\%    & 95.19\%  & 95.40\% & 94.82\% & 94.91\% & 95.34\% \\
        %         40\%    & 95.85\%  & 95.29\% & 94.56\% & 94.77\% & 94.70\% \\
        %         30\%    & 95.92\%  & 95.05\% & 95.87\% & 95.11\% & 96.55\% \\
        %         20\%    & 95.61\%  & 95.31\% & 95.71\% & 94.94\% & 97.30\% \\
        %         10\%    & 95.36\%  & 95.82\% & 96.30\% & 94.42\% & 91.02\% \\
        %         5\%     & 95.69\%  & 95.68\% & 93.43\% & 92.80\% & 90.98\% \\
        %         2.5\%   & 96.74\%  & 95.40\% & 91.93\% & 89.34\% & 85.52\% \\
        %         \botrule
        %     \end{tabular}
        % \end{table}

             \begin{table}[h]
            \caption{The 95\% credible interval coverage of all inferred parameters for five simulated data sets with reducing data retention levels.}\label{tab:95Levels}
            \centering
            \begin{tabular}{@{}l c c c c c@{}}
                \toprule
                Retention Level & Simulation 1 & Simulation 2 & Simulation 3 & Simulation 4 & Simulation 5 \\
                \midrule
                100\%   & 96.8\%  & 96.5\% & 95.4\% & 95.1\% & 96.0\% \\
                50\%    & 95.2\%  & 95.4\% & 94.8\% & 94.9\% & 95.3\% \\
                40\%    & 95.9\%  & 95.3\% & 94.6\% & 94.8\% & 94.7\% \\
                30\%    & 95.9\%  & 95.1\% & 95.9\% & 95.1\% & 96.6\% \\
                20\%    & 95.6\%  & 95.3\% & 95.7\% & 94.9\% & 97.3\% \\
                10\%    & 95.4\%  & 95.8\% & 96.3\% & 94.4\% & 91.0\% \\
                5\%     & 95.7\%  & 95.7\% & 93.4\% & 92.8\% & 91.0\% \\
                2.5\%   & 96.7\%  & 95.4\% & 91.9\% & 89.3\% & 85.5\% \\
                \botrule
            \end{tabular}
        \end{table}  

        % \textcolor{red}{There will be questions why there is no variability in the population estimates - justify. What about inference on other parameters- correlation/overdispersion/$\rho$? What can you conclude from the simulations? Is it good at estimating correlations/overdisperison etc? We see that the spatial surface is oversmoothed}

        Overall this simulation study illustrates that our framework is robust in the presence of declining and limited observation data, as is typically the case in species population studies. 
        While not unexpected, the oversmoothing of the underlying spatial surface (as seen in Figure~\ref{fig:SurfEstSims}) in the lowest data retention case is associated with higher RMSE values and wider 95\% credible intervals. 
        There is little variation in the mean estimates of the overall population of the three species across the different retention levels, however this is not the case at the most granular spatial level. 

        \subsubsection{Investigation on the Impacts of Sampling Regional Detection Probabilities}\label{subsubsec:VaryCullSims}
            We briefly examine the impact of sampling the regional detection probabilities (or cull percentages $\boldsymbol{\kappa}$) on the predicted total populations. 
            This is unlike Section~\ref{subsec:DeerSimResults} where $\boldsymbol{\kappa}$ was assumed constant and known. 
            The true underlying regional detection probabilities for each simulated data set is $\boldsymbol{\kappa} = 0.2$ for each species in each county. 
            This value was originally selected as it reflects scenarios within our motivating example. 
            We investigate the impact of sampling cull percentages when data retention is at its lowest, similar to our motivating example. 
            For illustrative purposes, we sample ten percentages from the distribution $\kappa \sim \mathcal{N}(\mu = 20, \sigma^2 = 2.55^2)$, similar to the interval outlined in Section~\ref{subsec:CullPercentagesUncert}. 
            That is, for this illustrative example, we examine the affect sampling regional detection probabilities $\boldsymbol{\kappa}$ when we retain 2.5\% of the original simulated data. 
            As outlined in Section~\ref{subsec:RegionalMod}, we cannot estimate both the regional detection probabilities, $\boldsymbol{\kappa}$, and the detection probabilities $\boldsymbol{p}$. 
            With the input of deer ecology experts, we can sample $\boldsymbol{\kappa}$ from an informative prior, and use these samples as input to our model. 
            
            % \textcolor{red}{Why 20\%? comment on poisson approx to binomial in regard to this}
            % \textcolor{red}{Also interested in estimating the correlation coefficients between species - you report this for deer? } 
            % We investigate when data retention is at its lowest, that is when we retain 2.5\% of the original simulated data. 
            % \textcolor{red}{As we cannot learn both the detection probabilities $\boldsymbol{p}$ and regional detection probabilities $\boldsymbol{\kappa}$ (YOU NEED TO INTRODUCE THIS IN ONE OF THE PREVIOUS MODELLING SECTIONS - DISCUSS THIS IDENTIFIABILITY PROBLEM IN MODELLING SECTION}. 
            % We reran the model with the new regional detection probabilities, and pool the generated samples. 

\begin{figure}[h]
                \centering
                \includegraphics[width = \textwidth]{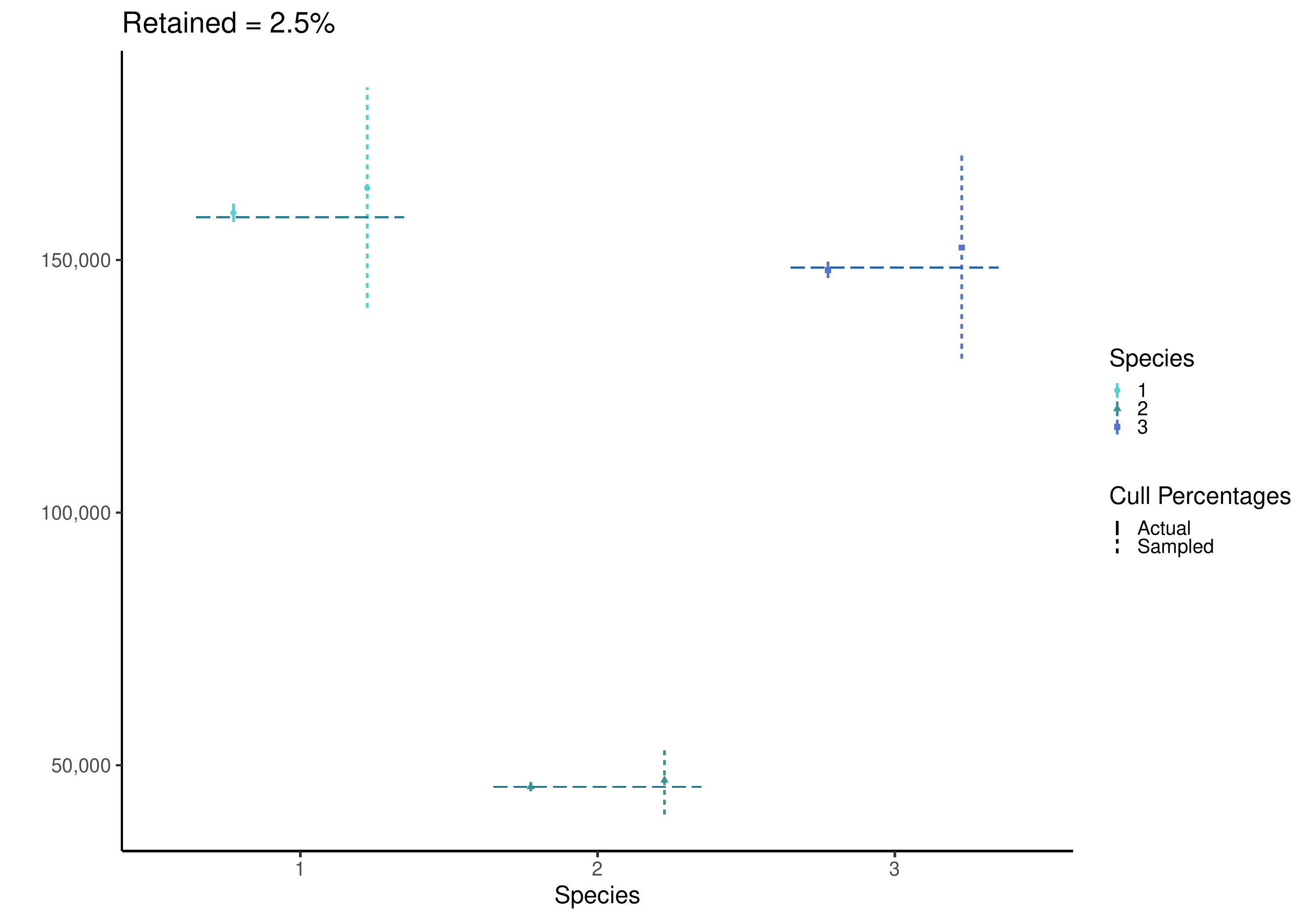}
                \caption{Impact of sampling 10 regional detection probabilities on 2.5\% of retained data from simulated data set 1.}
                \label{fig:SimCullSasmp}
            \end{figure}

            Figure~\ref{fig:SimCullSasmp} shows the additional variability in the population estimates when the regional detection probabilities are sampled. 
            The horizontal dotted lines indicates the true population for each species. 
            While the median estimates differ slightly between assuming the cull percentages are known and sampling these percentages, there is a substantial additional variability in the estimates produced when sampling has occurred. 
            
\begin{figure}[h]
                \centering
                \includegraphics[width=\textwidth]{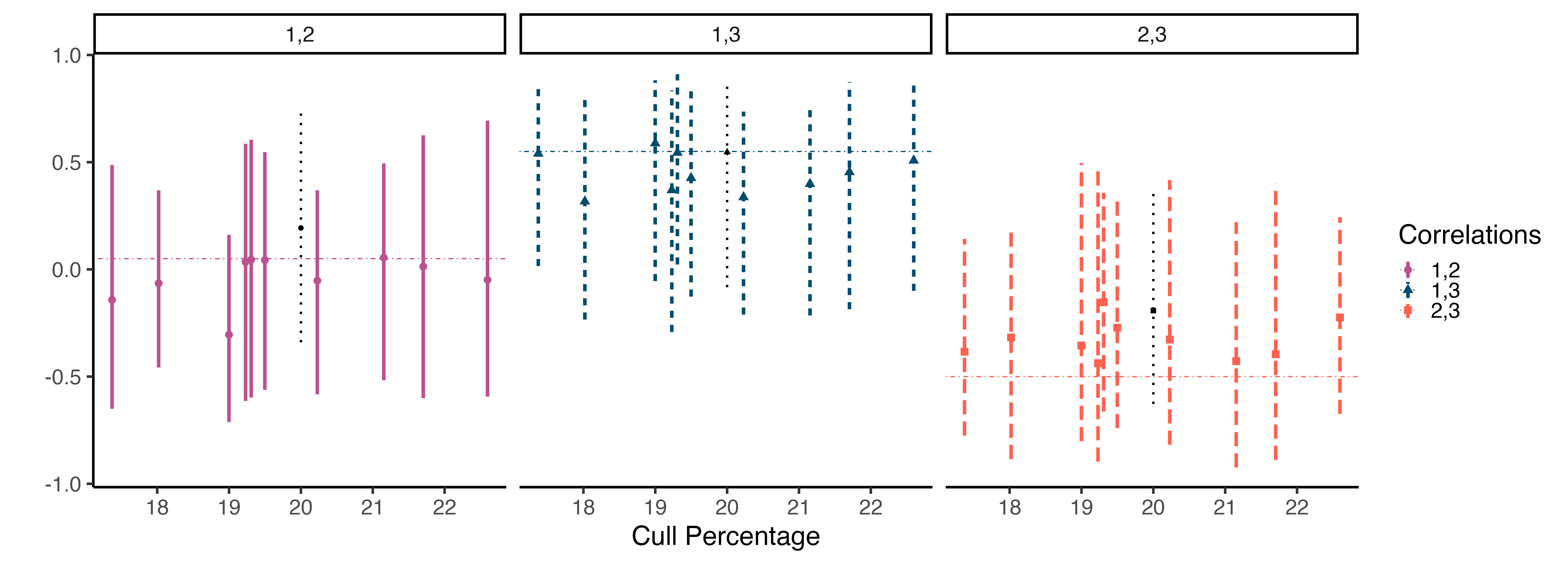}
                \caption{The impact of sampling cull percentages on the estimated between-species correlations. Shown in black is is the 95\% credible interval and mean estimate when the cull percentage is set to the correct underlying value of 20\%. The horizontal lines depict the true value for each.}
                \label{fig:sampCullCorr}
            \end{figure}

            We also assess the impact on the estimation of the between-species correlations. 
            In Figure~\ref{fig:sampCullCorr}, the mean estimate and associated 95\% credible interval for the correct regional detection probability is shown with a black filled circle and black dotted line respectively. 
            The width of these intervals remain similar across the sampled cull percentages, although the mean estimate of these correlations differ. 
           
\begin{figure}[h]
                \centering
                \includegraphics[width = \textwidth]{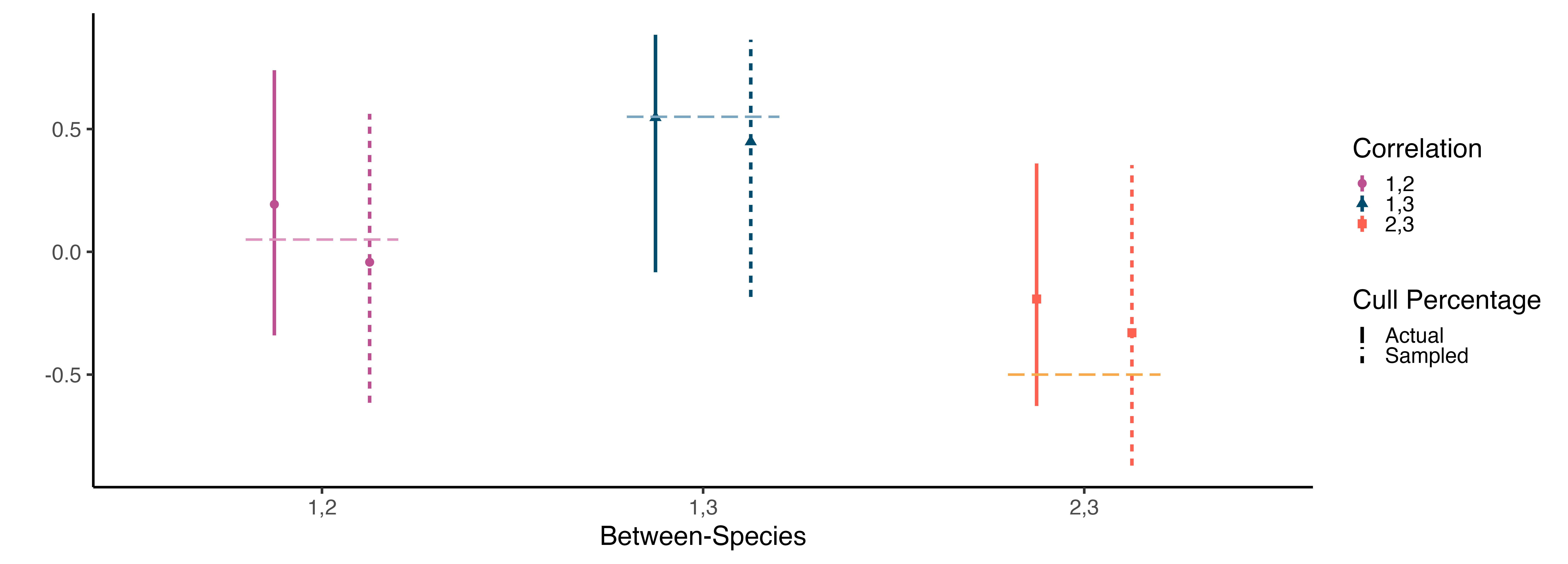}
                \caption{Comparison of sampling 10 cull percentages to estimate provided by taking the true underlying cull percentage with data retention at 2.5\%. The horizontal lines indicate the true correlation value.}
                \label{fig:AllCorsSim}
            \end{figure}

            However, when the 10 sampled cull percentages are pooled together, there is an increase in the variability of the estimates observed in Figure~\ref{fig:AllCorsSim} with respect to the situation where the cull percentages are known. 
            Figure~\ref{fig:AllCorsSim} shows a difference in the mean estimate for each of the between-species correlations when $\boldsymbol{\kappa}$ is sampled. 
            However, all three actual values are contained within the 95\% credible intervals in this scenario. 
            
%%%%%%%%%%%%%%%%%%%%%%%%%%%%%%%%%%%
     
\section{Case Study: Wild Deer in the Republic of Ireland} \label{sec:CaseStudy}

    For this problem, data regarding the number of hunting licences granted, the number of deer culled by recreational hunters and the number of deer observed was made available (sources: Irish Deer Commission; National Parks and Wildlife Service). 
    These data sets are spatially misaligned, with the number of licences granted and the recorded number of deer culled on a county or regional level, and the observed data across 732 squares at a 10km $\times$ 10km grid resolution. 
    We also incorporate proportion data on 14 different land types from the Corine land cover data set at the 10km $\times$ 10km resolution~\citep{CORINEDATASET.2018}. 
    
    In the Republic of Ireland, it is illegal to hunt deer without a licence. 
    Private deer hunting licences are granted by the National Parks and Wildlife Service (NPWS).
    Private licensed hunters are required to report the number of deer they culled in a season to the NPWS.
    This data is then collated to a county level and denotes both the overall number of culled deer and the breakdown by species. 
    Although these bagged numbers do not include all deer deaths (e.g. natural causes, poaching, road deaths), using these numbers has been shown to be a good predictor for deer populations~\citep{Adams:2020vv}.
    
\begin{figure}[h]
        \centering
        \includegraphics[width = \textwidth]{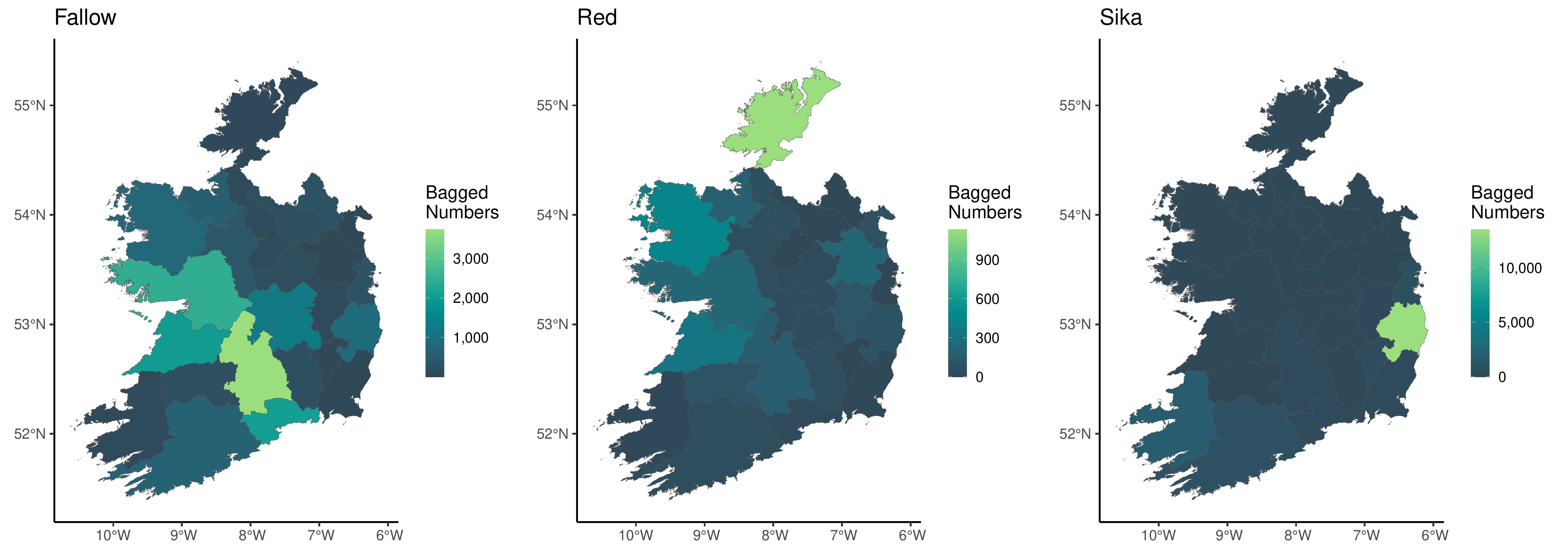}
        \caption{The number of deer bagged by hunters in each of the 26 counties of the Republic of Ireland during the 2017/2018 hunting.}
        \label{fig:Maps_Bagged}
    \end{figure}

    Figure~\ref{fig:Maps_Bagged} depicts the geographical nature of the data, showing the culled or bagged number of each species within the 26 counties of the Republic of Ireland during the 2017/2018 hunting season.
    During this hunting season, culls of red and sika deer were not recorded in every county (due to not being shot by hunters), at least 3 fallow deer were culled in every county.
    However, this does not preclude the absence of red and sika deer in these counties. 
    
\begin{figure}[h]
        \centering
        \includegraphics[width = \textwidth]{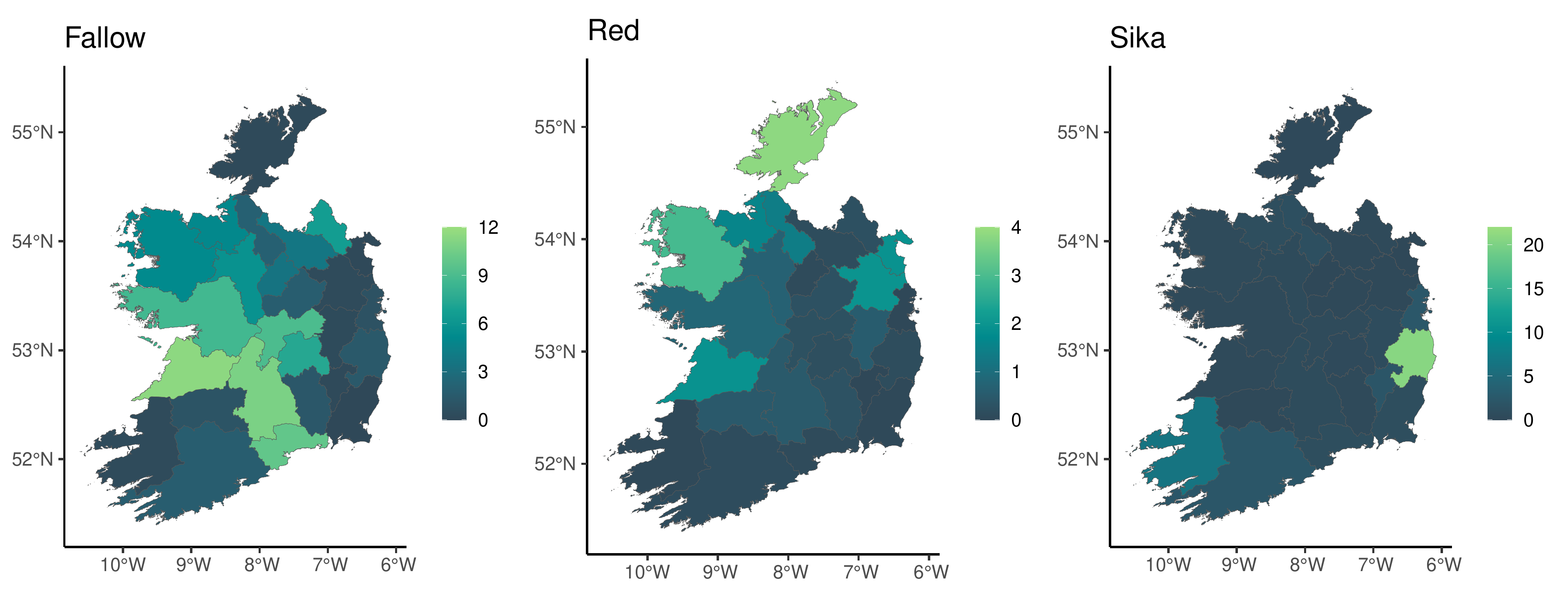}
        \caption{Bagged number of wild deer per licence for each species in the hunting season 2017/2018. Similar pattern seen across each of the three species as in Figure~\ref{fig:Maps_Bagged}.}
        \label{fig:HuntEffMap}
    \end{figure}

    For this analysis, we also accessed the number of granted deer hunting licences from the NPWS. 
    Combining these numbers with the bagged numbers, we calculated the number of culled deer per licence in each county. 
    This value may serve as a proxy for hunting effort, and combined with expert opinion, helps inform the ranges of our cull percentages. 
    Figure~\ref{fig:HuntEffMap} illustrates this proxy for hunting effort in each county of the Republic of Ireland for each individual deer species. 

\begin{figure}[h]
        \centering
        \includegraphics[width = \textwidth]{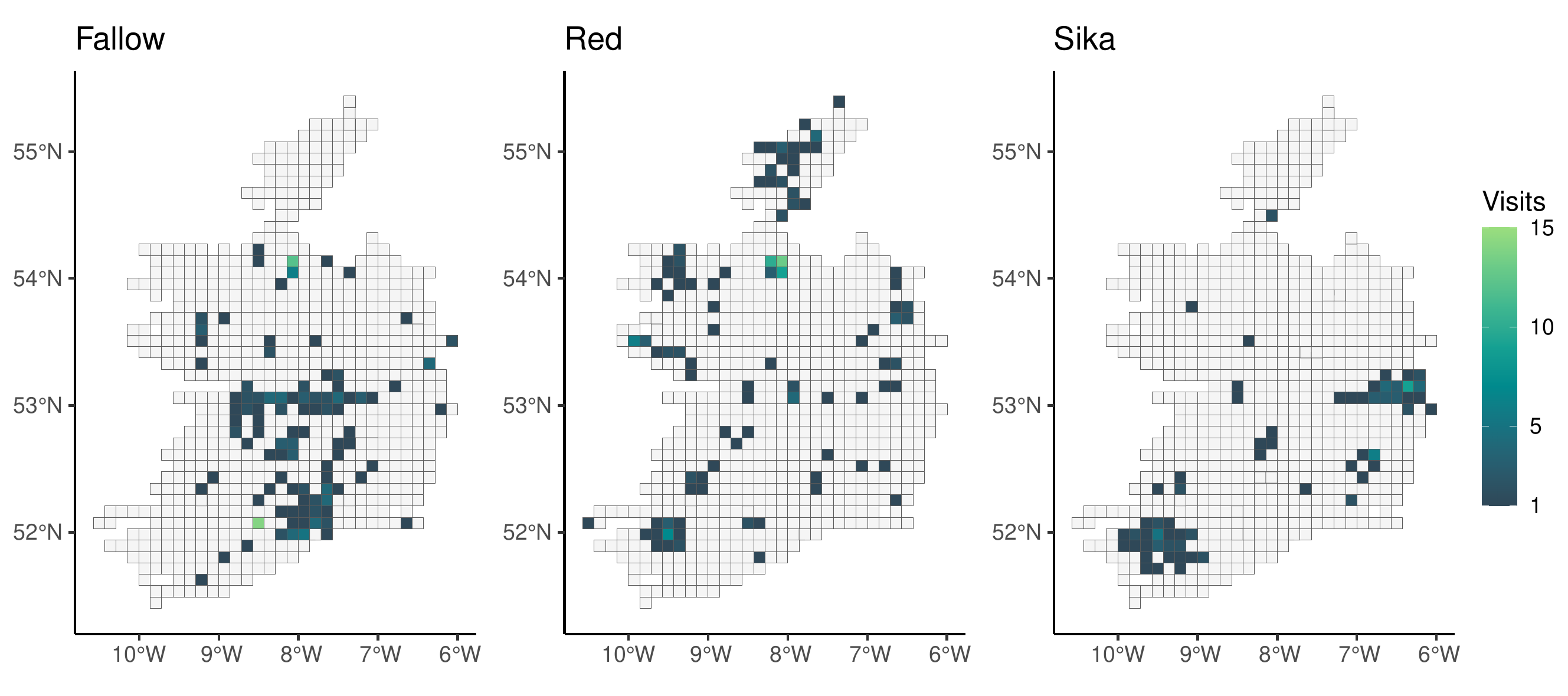}
        \caption{Squares coloured by the frequency of recorded visits.}
        \label{fig:maps_visits}
    \end{figure}

    During this hunting season, the Irish Deer Commission conducted a survey with hunters and have given us access to this data to conduct our analysis. 
    In this survey, they gathered data on sightings of the three main wild deer species at a finer spatial resolution, 10km $\times$ 10km grid level. 
    This data includes repeated site visits and repeated site visits where no deer were observed, resulting in presence-absence data and the ability to estimate detection probabilities. 
    The sampling effort (recorded number of visits) differs for each species. 
    Data on fallow deer has been accounted for in 96 of the 732 squares, totalling 184 site visits. 
    Red deer had data gathered in 91 squares which were visited 163 times. 
    There was little data gathered on the elusive sika deer, 105 visits in 61 squares. 
    Figure~\ref{fig:maps_visits} highlights the squares visited for each species and the number of times each square was visited. 
    In total, under 3\% of grid squares have observations.

    Our final data set is the 2018 Corine Land Cover (CLC)~\citep{CORINEDATASET.2018}.
    The CLC contains three class levels, each containing a different number of categories. 
    We selected, based on expert opinion, 14 different land types spanning the three class levels.
    We aggregate this data to a 10km $\times$ 10km resolution across the Republic of Ireland, recording the proportion of each land type within each of the 732 squares. 
    These 14 land types are amalgamated into 4 groups for the use in modelling the detection probabilities. 
    These groupings were informed on the experience of deer ecology experts and their informed knowledge from studying their natural habitat preference. 
    We model with the grouped land types to limit the effects of confounding in the model. 
    Table~\ref{tab:LandTypes} outlines the 14 different land types and the 4 groupings that will be used in our modelling. 

    \begin{table}[h]
        \caption{CLC classification of land cover variables used and the associated groupings of these variables used in modelling the detection probabilities.}
        \label{tab:LandTypes}
        \centering
        \begin{tabular}{@{}l l@{}}
            \toprule
            CLC Classification & Grouping \\
            \midrule
            Burnt Areas & Bad\\
            Urban Areas & Bad\\
            Agricultural Land & Agriculture \\
            Orchards & Good\\
            Pastures & Agriculture\\
            Broad Leaved Forests & Good\\
            Coniferous Forests & Coniferous and Peat\\
            Mixed Forests & Good\\
            Natural Grasslands & Agriculture \\
            Peat Bogs & Coniferous and Peat\\
            Transitionary Woodlands with Scrub & Good \\
            Beach & Bad\\
            Marsh & Bad\\
            Sea and Ocean & Bad\\
            \botrule
        \end{tabular}
    \end{table}

    \subsection{Model} \label{subsec:Model}
        As we have previously outlined the general framework in Section~\ref{sec:PropMod}, in this section we outline the modelling specifications for our motivating example. % This includes outlining our priors and the probability distributions for between- and within- site variation.

    %    For species $i$, let the $o_a$ observations be represented by $\mathbf{y}_a = \{y_{a,1}, y_{a,2}, \dots, y_{a, o_a} \}$. 
        % The unknown population of species $i$ at the desired grid scale, with $m$ areas, is denoted by $\mathbf{N}_i = \{N_{i1}, N_{i2}, \dots, N_{im} \}$. 
        % Then, for observation $k$ and area $j$,
        % \begin{equation*}
        %     \begin{split}
        %         n_ijk &\sim \text{Binomial} (N_{a, k \in n}, p_{a, k \in n}) \\
        %         N_{a,i} &\sim \text{Negative Binomial}(d_{a,i}, r_{a,i}).
        %     \end{split}
        % \end{equation*}
        %Allowing for overdispersion in the observed noisy counts, $\mathbf{N}_i$, we assume a negative binomial distribution for the population counts $\mathbf{N}_a$. 
        We adopt the parametisation of the negative binomial such that the mean abundance is $\boldsymbol{\lambda}$. 
        The remainder of the finer grid level model is given by
        \begin{equation}\label{eq:ActMod_P2}
            \begin{split}
                \text{logit}(p_{ij}) &= \delta_{0i} + G^\top_{j}\boldsymbol{\delta}_i \\
                \log(\lambda_{ij}) &= \beta_{0i} + X_{j}^\top\boldsymbol{\beta}_i + \phi_{i,j} \\
                \boldsymbol{\Phi} &= \Sigma \otimes \mathbf{u} \\
                \mathbf{u}_i &\sim \text{MVN}(0, [\tau_i(\mathbf{D} - \mathbf{A})]^{-1})\text{.}
            \end{split}
        \end{equation}
        The matrices $\mathbf{X}$ and $\mathbf{G}$ are matrices of of CLC land types and grouped land types, outlined in Table~\ref{tab:LandTypes}, respectively. 
        The multivariate correlated spatial surface is denoted by $\boldsymbol{\Phi}$, which we assign a multivariate intrinsic conditional autoregressive (MICAR) prior~\citep{Jin:2007uu}. 
        We assign an MICAR prior rather than other multivariate conditional autoregressive priors due to the estimation of one parameter per species, $\tau_a$, due to the paucity of the underlying data and computational burden. 
        % \textcolor{magenta}{We investigated using a MPCAR prior, with the addition of an additional parameter $\alpha$. However, the model suffered from poor mixing and did not converge, possibly due to the sparse nature of the data set. }

        There is a unit sum constraint on the rows of both covariate matrices, $\mathbf{X}$ and $\mathbf{G}$, as they contain proportion data. 
        While this an obvious and natural property, these matrices are not full rank which leads to difficulties for modelling purposes. 
        While log-ratio transformations have previously been used for compositional data, our data has meaningful zero values, thus these approaches are not applicable~\citep{Aitchison:1982vt, Aitchison:2005tq}. 
        To reduce the number of covariates and remove the unit sum constraint, we could apply principal component analysis to the matrices. 
        However, as one of our aims is for the model to be interpretable, we do not consider this approach. 
        We impose the Bayesian lasso penalty, proposed by Park and Casella~\cite{Park:2008uc}, on the coefficients of $\mathbf{X}$. 
        While other penalisation methods could be used, such as elastic net or ridge regression, we use the Bayesian lasso due to ease of implementation in \verb|nimble|.
        We do not impose a Bayesian lasso prior on $\boldsymbol{\delta}$ due to the small parameter space which was informed though expert opinion.

        At our county, or regional, level has $m = 26$ areal units. 
        For any county $j$, the population and culled number of deer of species $a$ is given by 
        \begin{equation} \label{eq:countyDeerMod}
            \begin{split}
                R_{ik} &= \sum_{j \in k} N_{ik} \\
                z_{ik} &\sim \text{Poisson}(\kappa_{ik}R_{ik})\text{.}
            \end{split}
        \end{equation}
        The population of species $i$ in county $k$ is given by $R_{ik}$, the cull percentages for species $i$ in county $k$ is represented by $\kappa_{ik}$, and the recorded number of bagged or cull deer of species $i$ is denoted by $z_{ik}$. 
        As previously mentioend, a Poisson approximation to the binomial distribution is used for the observed cull numbers to overcome mixing issues in the sampling chains observed for some model parameters. 
        % We use a Poisson approximation to the binomial distribution to evaluate the likelihood for the observed cull numbers to overcome mixing issues in the sampling chains observed in other model parameters. 
        For the sum in Equation~\ref{eq:countyDeerMod}, a grid square $j$ is assigned to county $k$ if the centroid of the square lies within county $j$.
        
        % Lastly, to ensure that the mean abundance of $N_{ij}$ is $\lambda_{ij}$, the parameters $d_{ij}$ and $r_{ij}$ of the multivariate process? \textcolor{red}{Update notation here in line with my changes - I don't understand what dij and rij are here?} are modelled as
        % \begin{equation*}
        %     \begin{split}
        %         d_{i,j} &= \frac{\nu_i}{\nu_i + \lambda_{ij}}        \\
        %         r_{ij} &= \frac{\lambda_{ij}d_{ij}}{1 - d_{ij}} \text{,}
        %     \end{split}
        % \end{equation*}
        % such that $\boldsymbol{\nu}$ controls overdispersion and the mean abundance of $\mathbf{N}$ is $\boldsymbol{\lambda}$. 
        Our remaining parameters for the fitted model are contained in the vector $\boldsymbol{\Theta} = \{\boldsymbol{\beta_0}, \boldsymbol{\beta}, \boldsymbol{\tau}, \boldsymbol{\delta}, \boldsymbol{\theta}\}$. 
        For the coefficients $\boldsymbol{\beta}$, we impose a Laplace(1,1) prior following that of the Bayesian lasso penalty to alleviate the effects of confounding between variables~\cite{Park:2008uc}. 
        The spatial precision parameter $\boldsymbol{\tau}$ is assigned a Gamma(shape = 10, scale = 0.5) prior, and $\boldsymbol{\theta}$ has an Exponential(0.5) prior assigned. 
        We assign a zero-mean Gaussian prior with a standard deviation of 5 to the coefficients of the detection probability $\boldsymbol{\delta}$ and species-specific intercept term $\boldsymbol{\beta_0}$. 
        The values for the components of the prior distributions were chosen after conducting simulation studies to assess their impact on parameter estimates. 
        
    \subsection{Addressing uncertainty in cull rates} \label{subsec:CullPercentagesUncert}
\begin{figure}[h]
            \centering
            \includegraphics[width = 14cm]{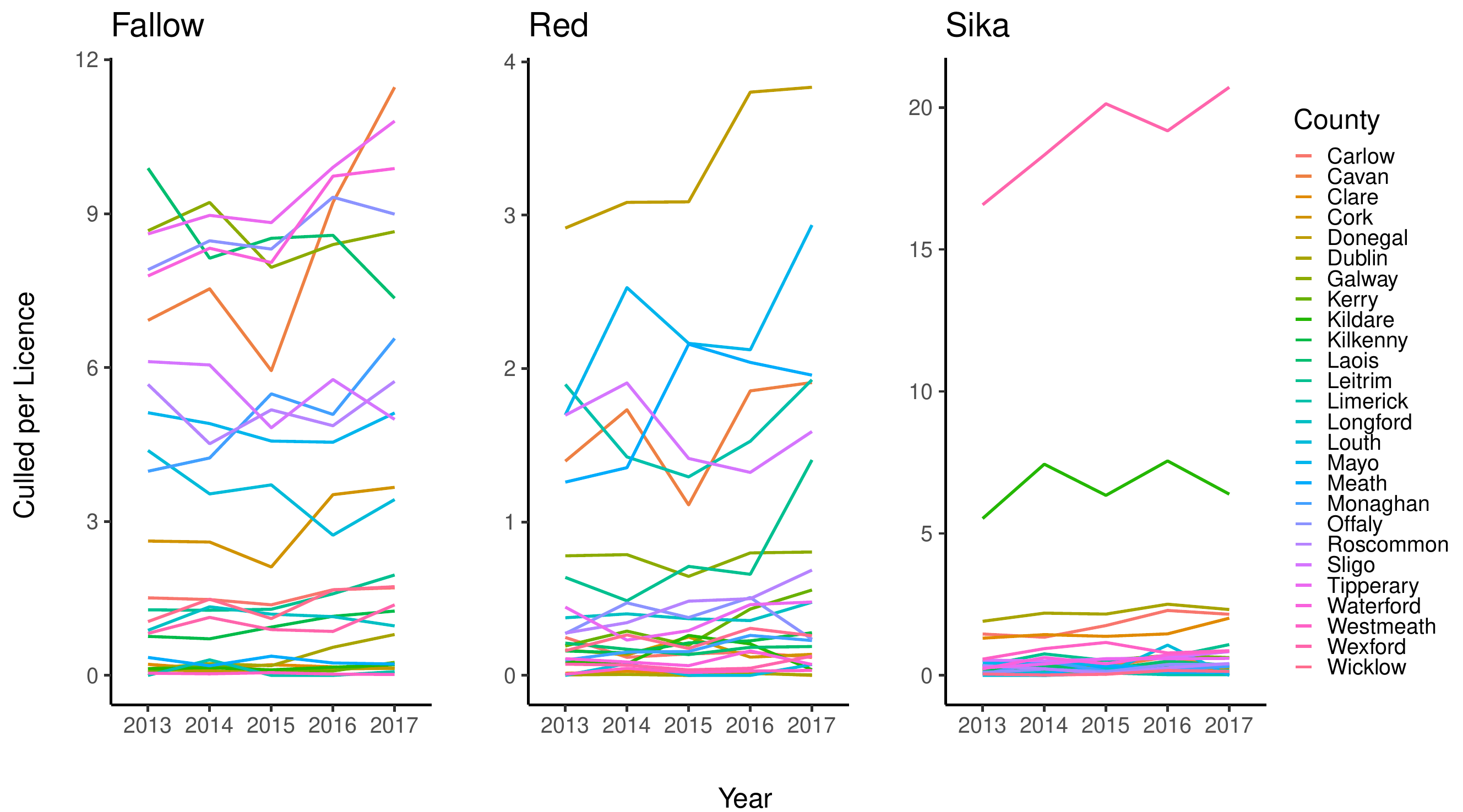}
            \caption{The number of culled deer per licence by species over 5 years.}
            \label{fig:BaggedPerLicence}
        \end{figure}

        Due to confounding and identifiability issues between detection probabilities at the 10km $\times$ 10km grid level and cull percentages at the county level, we cannot estimate both the detection probabilities and cull percentages. 
        We therefore sample cull values for each species in each county based on expert ecological opinion.
        We combine expert opinion and trends in the culled number of deer per licence to assign each county to an appropriate interval for each species.
        Figure~\ref{fig:BaggedPerLicence} shows these trends for each county by species. 
        Based on these trends, we classify each county into one of three bands, and these classifications may vary within each county by species. 
        The appropriate intervals for each of the three bands were suggested by deer ecologists, and the three bands are named \textit{Low} (between 5\% and 15\%), \textit{Mid} (between 10\% and 20\%), and \textit{High} (between 20\% and 30\%). 

        To translate the trends seen in Figure~\ref{fig:BaggedPerLicence} to the three bands, we need to consider the overall shape of the lines. 
        If a line does not decrease at any stage, that is the number of culled deer per licence has not decreased, that county is given a \textit{Low} classification. 
        The trend in County Donegal for red deer fits that of a county with a \textit{Low} cull percentage. 
        These counties are believed to have reproductive rates higher than the culling rate, thus allowing the increase in bagged numbers. 
        We sample these cull percentages from a $\mathcal{N}(\mu = 10, \sigma^2 = 2.55^2)$, roughly corresponding to a cull rate between 5\% and 15\%. 
        Counties where the reproductive rate and cull percentage are similar, thus maintaining the size of the herd, are grouped into the \textit{Mid} band, where cull rates are sampled from a $\mathcal{N}(15, 2.55^2)$, roughly corresponding to a cull rate between 10\% and 20\%. 
        Kerry is classified as having mid culling percentages for sika deer, as the starting and end points are roughly the same over the 5-year period.
        %as there is little variation between the start and end of the 5 year period. 
        Finally, counties where indications of reducing herd size are shown, are classified as counties with \textit{High} cull percentages, such that cull percentages are sampled from a $\mathcal{N}(25, 2.55^2)$, leading to cull rates between 20\% and 30\%. 
        A high culling percentage is only assigned for fallow deer in counties Laois, Longford, and Sligo, due to the decreasing trend seen across the 5 year period. 
        Table~\ref{tab:CullPrior} outlines for each county and species the assigned interval.
        \begin{table}[h]
            \caption{Classification for each cull percentage by county associated with the three wild deer species in the Republic of Ireland.}\label{tab:CullPrior}
            \centering
            \begin{tabular}{@{}l c c c@{}}
                \toprule
                County & Fallow & Red & Sika \\
                \midrule
                Carlow     & Mid  & Mid & Low \\
                    Cavan      & Low  & Mid & Mid \\
                    Clare      & Low  & Low & Mid \\
                    Cork       & Mid  & Mid & Low \\
                    Donegal    & Mid  & Low & Mid \\
                    Dublin     & Low  & Mid & Low \\
                    Galway     & Mid  & Mid & Low \\
                    Kerry      & Mid  & Mid & Mid \\
                    Kildare    & Mid  & Low & Low \\
                    Kilkenny   & Low  & Mid & Mid \\
                    Laois      & High & Mid & Mid \\
                    Leitrim    & Low  & Low & Mid \\
                    Limerick   & Mid  & Mid & Mid \\
                    Longford   & High & Mid & Mid \\
                    Louth      & Mid  & Mid & Mid \\
                    Mayo       & Mid  & Low & Mid \\
                    Meath      & Mid  & Low & Mid \\
                    Monaghan   & Low  & Mid & Low \\
                    Offaly     & Low  & Mid & Low \\
                    Roscommon  & Mid  & Low & Mid \\
                    Sligo      & High & Mid & Low \\
                    Tipperary  & Low  & Mid & Low \\
                    Waterford  & Low  & Mid & Low \\
                    Westmeath  & Low  & Mid & Mid \\
                    Wexford    & Mid & Mid & Low \\
                    Wicklow    & Low & Mid & Low \\
                \botrule
            \end{tabular}
        \end{table}

%%%%%%%%%%%%%%%%%%%%%%%%%%%%%%%%%%%%%%%%%%%%%%%%%%%%%%%%%%%%%%
%%%%%%%%%%%%% ALL RESULTS, BOTH SIMS AND ACTUAL %%%%%%%%%%%%%%
%%%%%%%%%%%%%%%%%%%%%%%%%%%%%%%%%%%%%%%%%%%%%%%%%%%%%%%%%%%%%%
    \subsection{Results} \label{sec:Results}\label{subsec:DeerResults}
    % \textcolor{brightOrange}{All analysis was conducted using a DELL XPS 15 9570 laptop with 6 2.9 GHz Dual-Core Intel i9-8950 HK processors and 32 GB of memory. 
    % All computation times recorded are for models run using this laptop. 
    % }
    %\subsection{Case Study}
    %\label{subsec:DeerResults}
        This section interprets the results of running the model with 200 different cull scenarios, with culling percentages sampled from the intervals previously outlined in Section~\ref{subsec:CullPercentagesUncert} and Table~\ref{tab:CullPrior}. 
        We generate 200 cull scenarios, as from previous simulation studies the credible intervals for the population estimates do not change to any significant degree with the consideration of additional cull scenarios.
        We further examine the effects of sampling 200 cull percentages in Section~\ref{subsec:CullsSamp}. 
        Each scenario contains a cull percentage for each of the 26 counties and each of the three species, resulting in 78 (26 $\times$ 3) different cull percentages per scenario.
        Unlike the work conducted by Kelly et al.~\cite{Kelly:2021wz}, in which only one fixed cull percentage is assumed at the county level, sampling the cull percentages adds additional uncertainty in population estimates, reflective of the uncertainty in the cull percentages themselves. 
        For each scenario, we generated 1,000 independent samples from the posterior distribution for each model parameter, resulting in our analysis being conducted on 200,000 samples from the posterior distribution in total.
        These samples were generated using 10 chains run in parallel, each generating 100 independent samples post burn-in, taking approximately 80 minutes to run. 
        Posterior checks were carried out to ensure independence of the samples and sufficient burn-in. 
        This model, with 200 cull scenarios, took approximately 11 days using a DELL XPS 15 9570 laptop with 6 2.9 GHz Dual-Core Intel i9-8950 HK processors and 32 GB of memory.
        % The total run-time for the analysis was approximately 11 days .

\begin{figure}[h]
            \centering
            \includegraphics[width = 14cm]{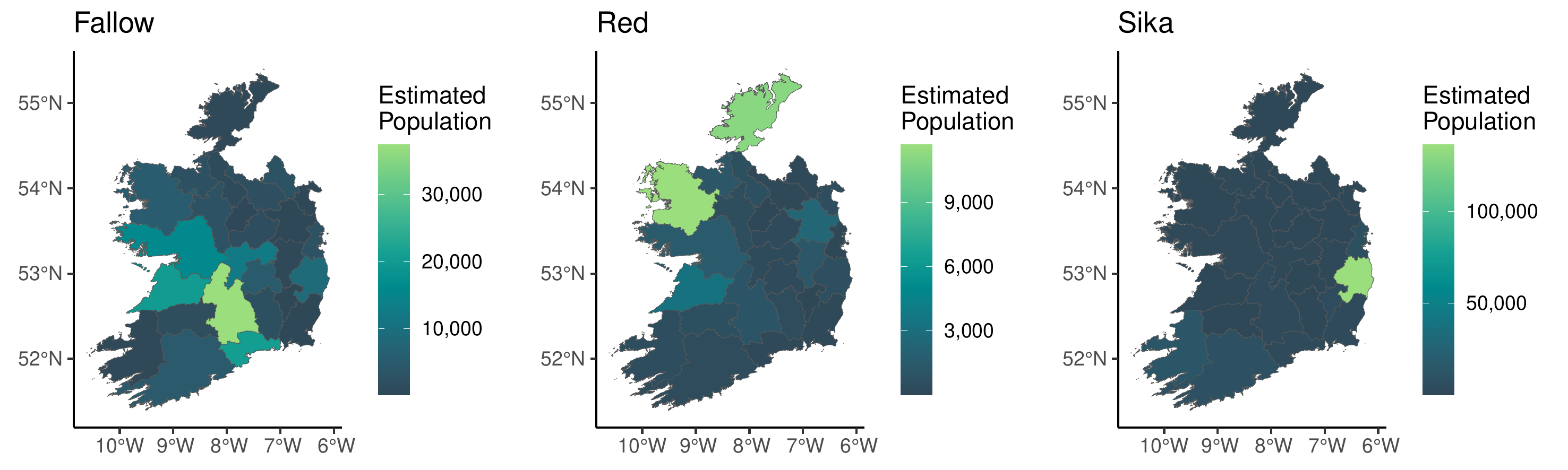}
            \caption{Estimated median populations for each deer species shown on a map of the Republic of Ireland.}
            \label{fig:EstCountyMap}
        \end{figure}

        Table~\ref{tab:CountyPops} shows the estimated species populations by county for the Republic of Ireland, and estimated total populations. 
        Across the Republic of Ireland, sika deer are estimated to be the most populous species of deer (median estimate of 185,056), with the median estimated population of fallow deer lower at 155,964.
        Unsurprisingly, the native red deer have the smallest population with a median estimate of 39,211. 
        Red deer almost disappeared in the 20$^{\text{th}}$ century, but due to rigorous protection and management, their population has increased~\citep{Carden:2012wo}. 
        Figure~\ref{fig:EstCountyMap} illustrates the estimated median population for each species by county.
        
        \begin{table}[h]
            \caption{Population for each species of deer given by county and across the Republic of Ireland. Included in the table are the estimated median and the associated 95\% credible intervals.}\label{tab:CountyPops}
            \centering
            \begin{tabular}{@{}l l l l@{}}
                \toprule
                County & Fallow & Red & Sika \\
                \midrule
                    Carlow     &     175 [106 - 297]        & 11 [1 - 36]               & 4,051 [2,591 - 9,100] \\
                    Cavan      &   1,862 [1,197 - 4,331]    & 61 [29 - 119]             & 40 [17 - 84] \\
                    Clare      &  20,349 [13,900 - 34,128]  & 3,289 [2,171 - 7,063]     & 322 [218 - 522] \\
                    Cork       &   4,552 [3,445 - 6,992]    & 340 [225 - 532]           & 8,215 [5,565 - 15,314] \\
                    Donegal    &     268 [176 - 428]        & 11,037 [7,806 - 22,851]   & 425 [289 - 705] \\
                    Dublin     &   2,614 [1,650 - 4,947]    & 21 [5 - 55]               & 7,495 [5,156 - 16,306] \\
                    Galway     &  15,873 [12,059 - 23,085]  & 1,490 [1,072 - 2,275]     & 872 [543 - 1,565] \\
                    Kerry      &     414 [218 - 640]        & 428 [365 - 524]           & 13,036 [9,49970 - 18,737] \\
                    Kildare    &     451 [330 - 641]        & 1,191 [726 - 2,275]       & 1,302 [855 - 2,786] \\
                    Kilkenny   &   2,080 [1,362 - 3,959]    & 305 [202 - 487]           & 398 [278 - 637] \\
                    Laois      &   4,974 [4,173 - 6,157]    & 225 [147 - 356]           & 425 [291 - 637] \\
                    Leitrim    &     959 [608 - 1,833]      & 642 [410 - 1,429]         & 333 [219 - 534] \\
                    Limerick   &   1,302 [950 - 2,100]      & 620 [438 - 978]           & 251 [165 - 404] \\
                    Longford   &     225 [169 - 306]        & 14 [3 - 43]               & 9 [1 - 29] \\
                    Louth      &      99 [68 - 147]         & 521 [362 - 780]           & 10 [2 - 33] \\
                    Mayo       &   5,272 [3,976 - 7,649]    & 11,665 [9,917 - 15,681]   & 390 [260 - 623] \\
                    Meath      &     188 [119 - 319]        & 2,328 [1,494 - 4,709]     & 88 [47 - 162] \\
                    Monaghan   &   2,804 [1,924 - 6,369]    & 68 [33 - 131]             & 129 [61 - 312] \\
                    Offaly     &  12,405 [8,103 - 23,123]   & 234 [148 - 372]           & 537 [328 - 1,239] \\
                    Roscommon  &   2,562 [1,902 - 3,788]    & 454 [267 - 1,007]         & 194 [122 - 328] \\
                    Sligo      &   2,348 [1,882 - 2,927]    & 1,238 [865 - 1,759]       & 990 [641 - 1,881] \\
                    Tipperary  &  37,358 [24,705 - 78,751]  & 1,083 [779, 1,548]        & 2,037 [1,328 -4,348] \\
                    Waterford  &  20,692 [14,011 - 38,657]  & 107 [59, 193]             & 1,806 [1,131 - 3,602] \\
                    Westmeath  &   1,427 [921 - 3,393]      & 149 [90 - 250]            & 80 [42 - 145] \\
                    Wexford    &     109 [71 - 174]         & 91 [58 - 150]             & 2,537 [1,542 - 5,302] \\
                    Wicklow    &   8,791 [5,557 - 16,335]   & 541 [371 - 875]           & 136,130 [90,023 - 252,436] \\
                    \textbf{Total}      & \textbf{155,964 [133,669 - 213,053]} & \textbf{39,211 [34,033 - 52,264]} & \textbf{185,056 [135,793 - 298,827]} \\
                \botrule
            \end{tabular}
        \end{table}
        
        Overall, the estimated population numbers from the model are aligned to what we know about presence distribution of wild deer and anecdotal observations from deer hunters, farmers, and conservation personnel, and reports on damage caused by over localised populations of wild deer in many areas to various land uses and conservation habitats and vehicular accidents involving deer on many primary roads and secondary roads. 
 
\begin{figure}[h]
            \centering
            \includegraphics[width = \textwidth]{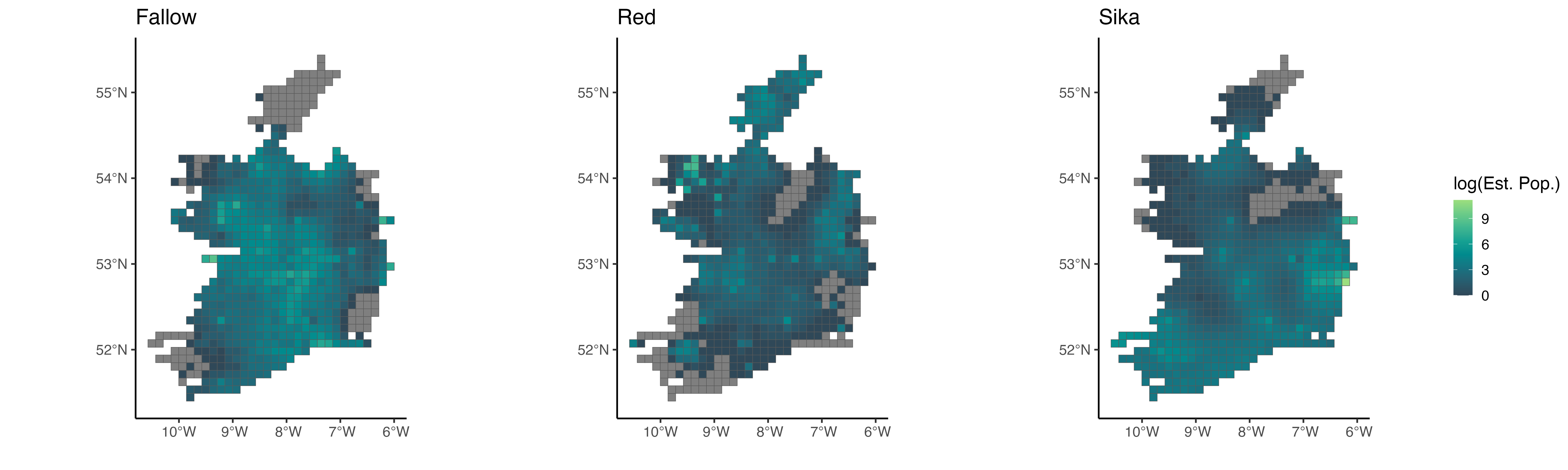}
            \caption{Log median estimated population at each grid square. Grey indicates squares where the median estimated population was 0.}
            \label{fig:squarePopLog}
        \end{figure}

        Deer population estimates are also available at the 10km $\times$ 10km grid level.  Figure~\ref{fig:squarePopLog} is shown on the $\log$ scale, in order to examine the patterns formed for each species. 
        These maps align with the estimated median county population maps shown in Figure~\ref{fig:EstCountyMap}. 
        The patterns observed for each species is in-line with previous literature~\citep{CARDEN:2011vm, Morera-Pujol:2023wj, Murphy:2023uh}, known localised hot-spots for individual species, and the known expansion rates of the three deer species.
        
        Fallow deer are the most widely distributed species in the Republic of Ireland, as observed in Figure~\ref{fig:EstCountyMap}, and there has been little change since the 2008 survey by Carden et al.~\cite{CARDEN:2011vm}, although they are expected to increase their total range distribution at relatively low rates (3\% compound annual rate of expansion).
        However, the herd may build up to high densities but remain in the area of their original release, thus creating large populations beyond the carrying capacity of the habitats. 
        The landscape of the largely limestone-based Irish lowlands is preferred and well suited to fallow deer.
        The highest concentration of estimated numbers the model produced was in County Tipperary, which is a county based on limestone and of enriched agricultural crops, including grasslands for domestic livestock grazing and broadleaf forestry/woodlands.
        The estimates presented here follow similar patterns of presence as per Carden et al.~\cite{CARDEN:2011vm}, therefore indicating preferred habitats for foraging and reproduction has not changed much in the previous 10 years.
        However, they have expanded their range to the eastern midlands and this may be due to higher densities being reached in areas.  
        This range expansion for fallow deer can also be seen in Murphy et al.~\cite{Murphy:2023uh}. 
        
        Although red deer can expand their range at high rates (7\% annually over a 30-year period~\citep{CARDEN:2011vm}), red deer are the least widely distributed species, being limited mainly to County Kerry (Killarney National Park and immediate surrounds), counties Galway and Mayo, and Glenveagh National Park in County Donegal. 
        There are smaller outlier populations of red deer to the known distribution hot spots across Ireland.
        However, there are no red deer left in County Wicklow due to hybridisation with sika since 2008. 
        Red deer have a natural preference for deciduous broadleaf woodlands, but have been pushed out into more open upland habitats by humans, thus limiting their population through this ecological adaptation across the mountain regions. 
        In the upland peat areas in the west and northwest of the Republic of Ireland, there are large commercial plantations of Sitka spruce, where red deer find shelter and forage. 
        They may negatively impact on the trees through browsing of lateral and leader shoots of saplings and immature trees and bark stripping of semi-mature trees.
        
        It has been shown in a previous 30-year study that sika in Ireland expand their total range by 5\% per annum~\citep{CARDEN:2011vm}.
        However, by far the most human-deer conflict issues with sika occur in the east of the country in County Wicklow, where anecdotes and evidence show they are causing damage to conservation habitats, commercial plantation coniferous and broadleaf forestry and to agricultural crops. 
        It’s unsurprising that the model has identified County Wicklow as it is a hot spot of high numbers of sika.
        There are large tracts of uplands, which sika have adapted to, of peat bog and mountains with large scale Sitka/other commercial forestry plantations in which they use for shelter and foraging, as well foraging on agricultural crop lands and grasslands adjacent to the plantations. 
        Through much of County Wicklow, due to a large issue of illegal shooting at night under a lamp, or even in daylight hours, (poaching), the deer have adapted a more nocturnal lifestyle, emerging from dense forest to feed in dark hours, and thus managing these herds by the recreational hunters during legal shooting daylight hours has become quite difficult, if impossible in many areas. 
        A combination of dense habitats, inaccessible uplands, the presence of a national park where recreational hunters cannot shoot in, and the opportunistic ecological behaviour of sika may have led to the overabundance of sika numbers in County Wicklow relative to other counties where sika are found in much smaller estimated population numbers.

\begin{figure}[h!]
            \centering
            \includegraphics[width=14cm]{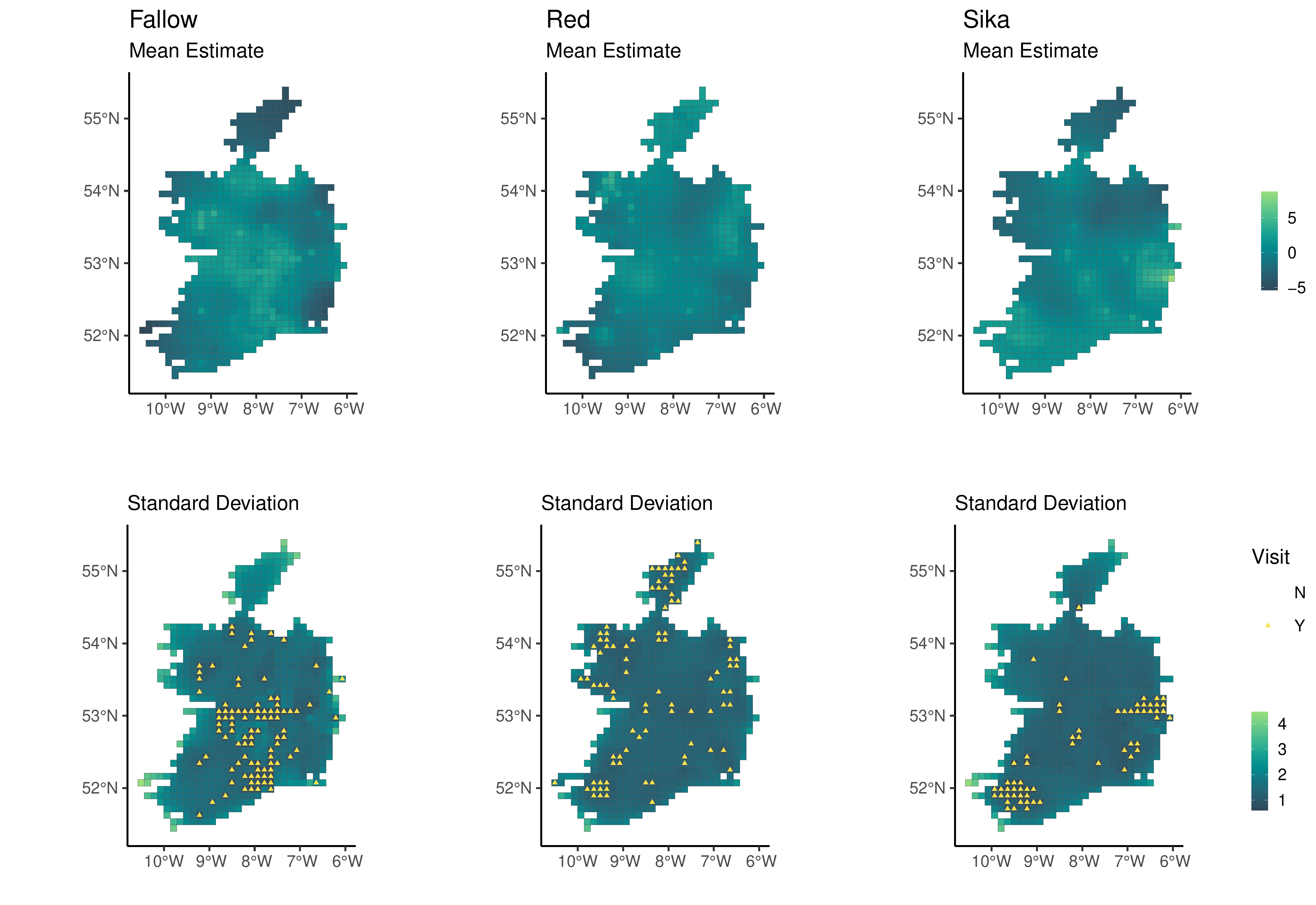}
            \caption{Estimated mean spatial surface for each wild species of deer and the associated standard deviations at each grid square. The yellow triangles on the standard deviation map indicates grid squares where we have at least one recording from our observational data. }
            \label{fig:Surfaces}
        \end{figure}

        Figure~\ref{fig:Surfaces} illustrates the spatial surface for each species from the MICAR component $\boldsymbol{\phi}$. The top row of Figure~\ref{fig:Surfaces} shows the mean estimate for each surface, and the lower row is the associated standard deviation. The additional markings on the standard deviation maps indicate squares that have had visits. 
        The pattern seen across all three species is similar to the presence and relative-abundance maps produced by Morera-Pujol et al.~\cite{Morera-Pujol:2023wj}, relative-density maps produced in Murphy et al.~\cite{Murphy:2023uh}, and the binary species distribution maps seen in Carden et al.~\cite{CARDEN:2011vm}. 
        The hot-spots seen for all three species also align with expert opinion, in particular the known hot spots for sika in the east, southwest and northwest, with smaller population outliers elsewhere.
        These patterns also align with the log median population estimates at this grid level shown in Figure~\ref{fig:squarePopLog}.

\begin{figure}[h!]
            \centering
            \includegraphics[width = \textwidth]{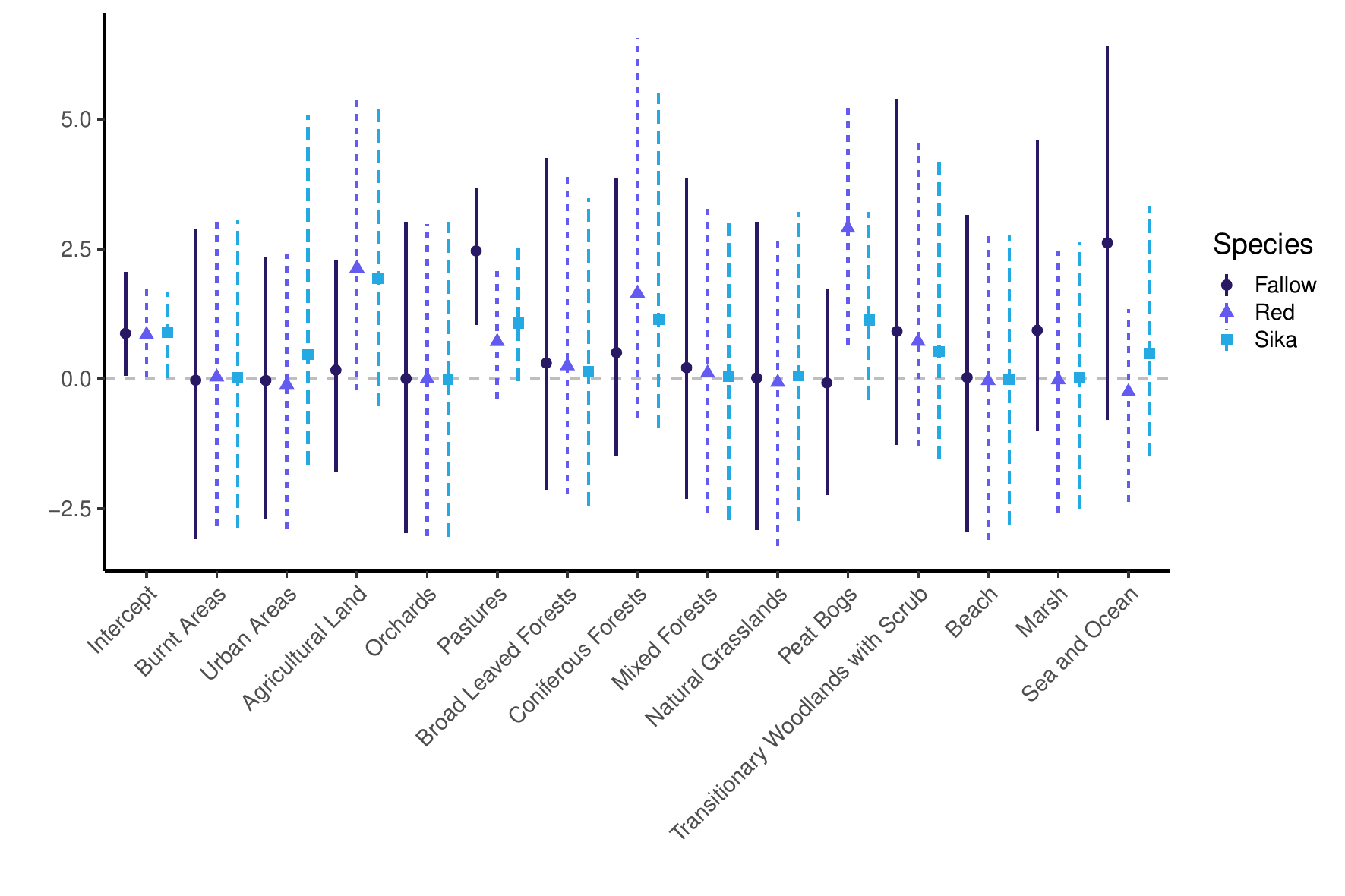}
            \caption{Estimated coefficients for land covering for each species. }
            \label{fig:coeffsBetas}
        \end{figure}

        In Figure~\ref{fig:coeffsBetas} the estimates for each coefficient of the 14 chosen CLC land coverings, along with the associated 95\% credible intervals are presented. 
        We observe that of the 42 estimated $\beta$'s associated with the CLC land types, only two credible intervals do not contain zero. 
        Unsurprisingly, \textit{Pastures} is positive for fallow deer. 
        Fallow deer are a lowland species, preferring a mosaic of open woodland and pasture to shelter and forage within.
        Due to the forced migration of red deer to the uplands and bog lands across the Republic of Ireland, ecologically the positive association between red deer and \textit{Peat Bogs} is tenable.
        Due to the sparse nature of the data, there is a large degree of uncertainty in the parameter estimates for covariates. 
        Given additional observational data for each species, we would expect the credible intervals in Figure~\ref{fig:coeffsBetas} to increase in precision. 
        We observe large, positive, median estimates for the regression parameters corresponding to \textit{Agricultural Land}, \textit{Pastures} and \textit{Coniferous Forests} for red and sika populations. 
        Based on expert ecological opinion, these correspond to the preferred feeding habitats of these species as well as reflecting the migration of these species to areas of lower human habitation and uplands as represented  by the percentage of coniferous forest plantations. 
        The large positive coefficient for \textit{Sea and Ocean} observed for fallow is an artefact of a small population based on Lambay island off the coast of County Dublin. 

\begin{figure}[h!]
            \centering
            \includegraphics[width = \textwidth]{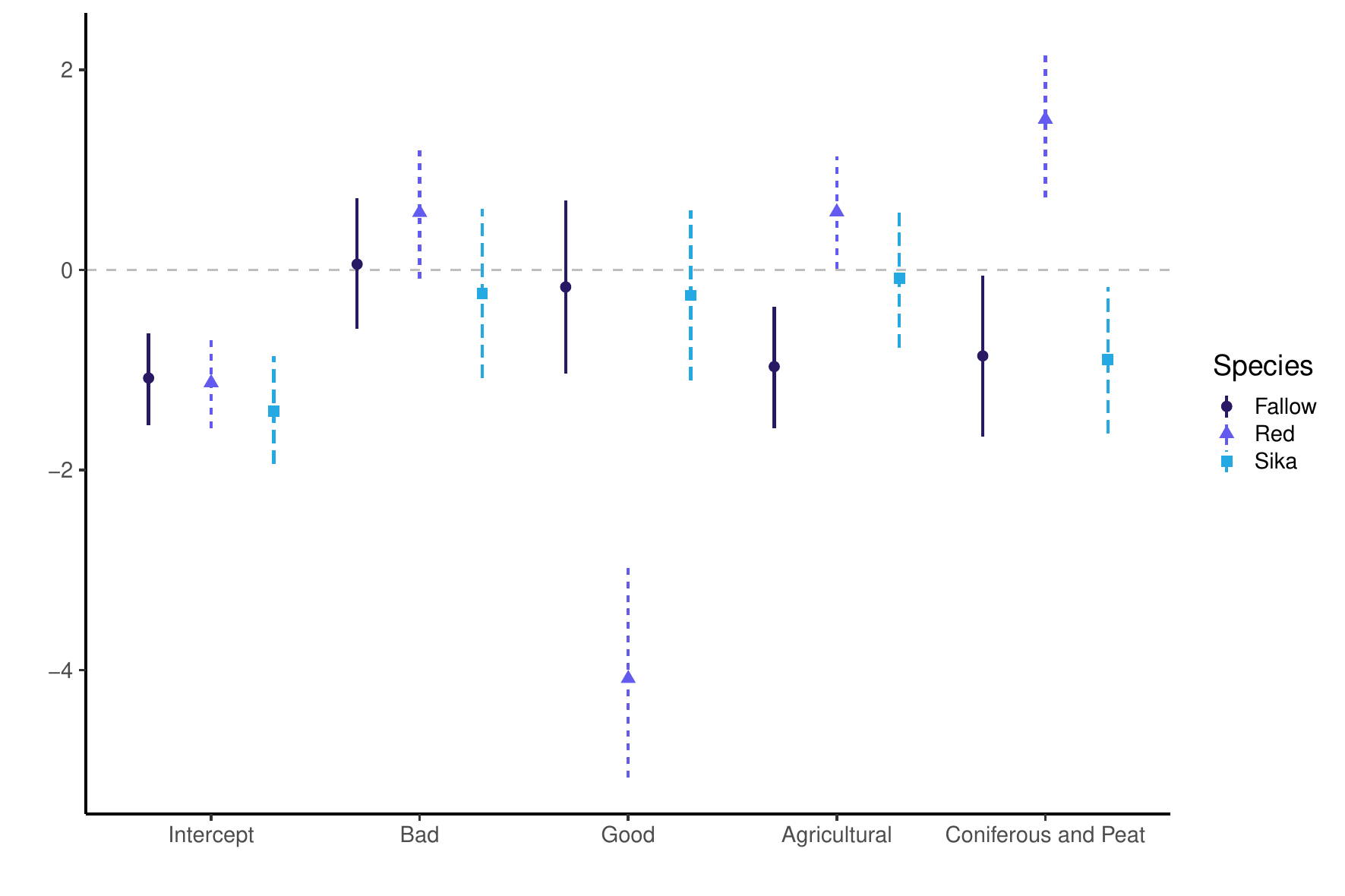}
            \caption{Coefficients for detection probabilities. The lower the value of $\delta$, the smaller the probability of detection.}
            \label{fig:Deltas}
        \end{figure}

        Figure~\ref{fig:Deltas} shows the estimated values for the coefficients for the grouped land coverings that inform our detection probabilities.
        The impact of a lower covariate coefficient is shown when the percentage of the associated land grouping is high, by lowering the associated probability of detection. 
        The four groups are classified as \textit{Bad}, \textit{Good}, \textit{Agriculture}, and \textit{Coniferous and Peat} (as shown in Table~\ref{tab:LandTypes}), while also including an intercept term. 
        The \textit{Good} coefficient for red deer is quite negative, suggesting that the higher the percentage of good land in a square, the lower the probability of detection of a red deer is for a given site visit, provided they are present. 
        This is presumably a reflection of the small numbers of red deer with habitats of this type. 
        Conversely, the \textit{Coniferous and Peat} coefficient for red deer is positive, indicating that the higher the percentage of coniferous forests and peat bogs in a square, the higher the probability of detecting a red deer is on a given site visit. 
        This is presumably a reflection of their grazing habits - conversely the negative coefficient for sika deer is an artefact of changing grazing patterns given hunting and poaching pressures.
        There may be an element of slight statistical confounding between the estimated abundance and presence parameters, although substantially more data would be required to establish this.  
        % \textcolor{red}{ Perhaps a  comment along the lines of Gelman and Hill about the sign of the covariate making sense (even if not statistically significant) and so we leave it in? }

\section{Sensitivity of Population Estimates to Cull Parameters} \label{sec:SensitivityToCulls}
        In this section we explore the impact of the selected cull percentage intervals on our results. 
        We reran our model using only the mean cull percentage and compare the results to those seen in Section~\ref{subsec:DeerResults}, similar to work previously conducted in Ireland~\citep{Kelly:2021wz}.
        Section~\ref{subsec:ChangeCull} shows the impact  of changing the assumed cull percentage in one county on the sika population results. 
        We provide some rational into sampling 200 cull percentages in Section~\ref{subsec:CullsSamp}.
        We provide an ecological perspective and validate the estimated populations in Section~\ref{subsec:EcologicalPerspective}. 
        
        \subsection{Sensitivity of Results to Ignoring Cull Uncertainty at the County Level} \label{subsec:SampleVMean}
            A previous study used a constant cull percentage when estimating sika deer populations in County Wicklow~\citep{Kelly:2021wz}.
            We explore the impact failing to account for the uncertainty in cull percentages on the population estimates, with the results shown in Table~\ref{tab:PopEstimatesComp}. 
            While the median estimates are similar for the three species, there is a dramatic difference in the associated 95\% credible intervals. 
            The range of values covered is considerably larger when accounting for uncertainty in the cull percentages, with a significant increase in the upper end of the interval.
            This increase in range is due to the additional uncertainty added from sampling different cull percentages. 
            
            \begin{table}[h]
            \caption{Comparison of median wild deer population estimates and associated 95\% credible interval for 200 sampled cull rates and mean cull rate.}
            \label{tab:PopEstimatesComp}
            \centering
            \begin{tabular}{@{}c c c c@{}}
                \toprule
                Cull percentages & Fallow & Red & Sika \\
                \midrule
                200 Sampled & 155,964 [133,669 - 213,054] & 39,211 [34,033 - 52,264] & 185,056 [135,793 - 298,827] \\
                Mean & 150,384 [147,956 - 153,505] & 38,444 [37,339 - 39,555] & 180,706 [178,010 - 183,883] \\
                \botrule
            \end{tabular}
        \end{table}

\begin{figure}[h]
            \centering
            \includegraphics[width=\textwidth]{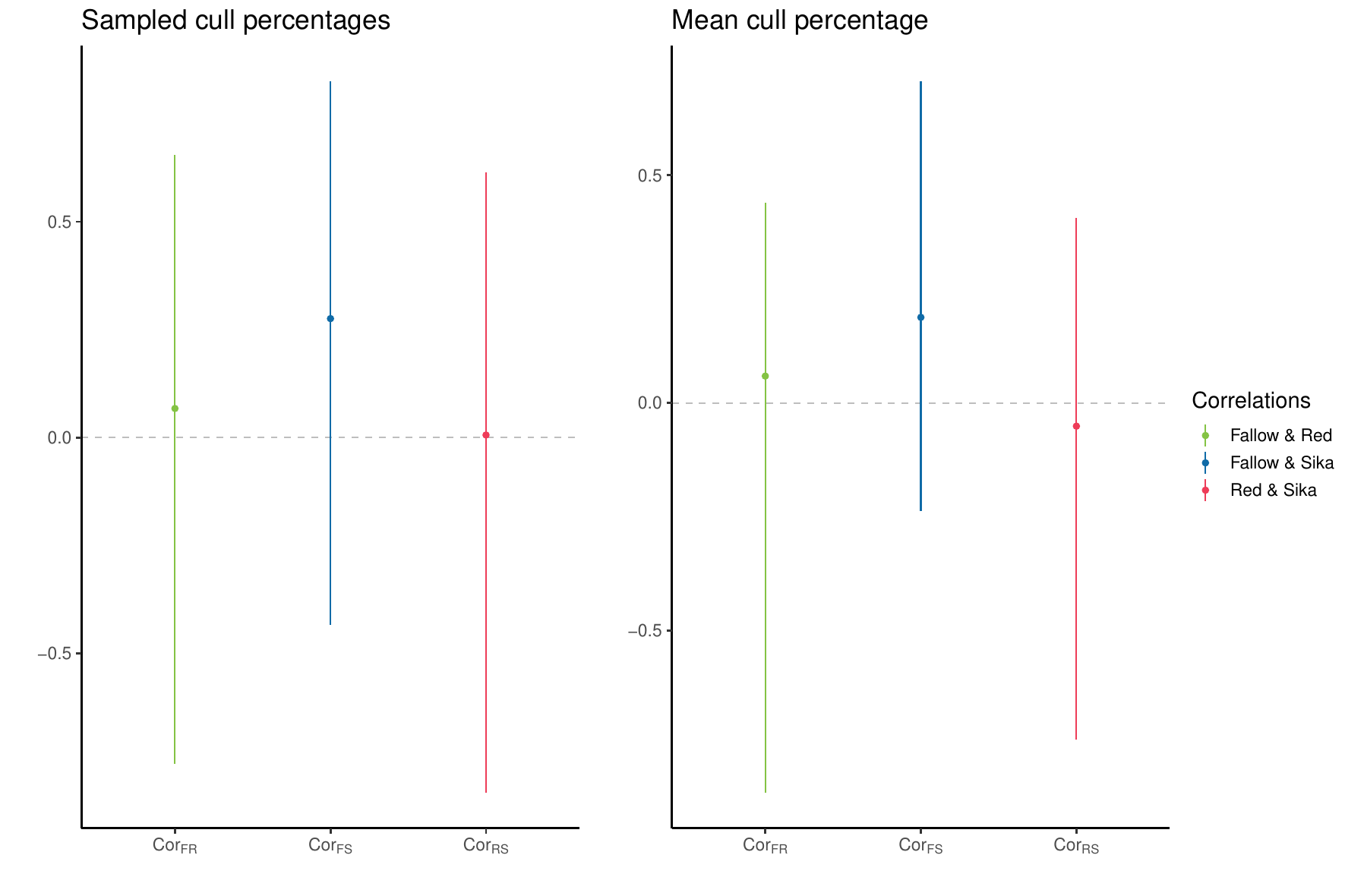}
            \caption{Estimated between-species correlations, comparing between 200 sampled cull percentages and using the mean cull percentages.}
            \label{fig:Correlations}
        \end{figure}

        It is also interesting to look at the impact sampling cull percentages has on other parameters. 
        Due to the formulation of the correlated spatial surface, we can calculate the residual between-species correlation included in the surface. 
        These correlation estimates account for the unexplained or left over correlation, after we take all other variables in the model, such as land type, into account. 
        Figure~\ref{fig:Correlations} shows the estimates and associated 95\% credible intervals for the three between-species correlations for both the 200 sampled cull percentages and when we ignore cull uncertainty.
        The variability differences can clearly be seen in the two plots, with narrower credible intervals when cull uncertainty is ignored.
        However, there is a great deal of uncertainty in these estimates due to the low quality of data available for model fitting. 
                
    \subsection{Sensitivity to Choice of Cull Ranges} \label{subsec:ChangeCull}
        Sika deer were originally introduced to the Republic of Ireland in 1860 to the Powerscourt Estate in County Wicklow~\citep{Powerscourt:1884wc}. 
        From the NPWS hunting returns for 2017/18 deer season, a total of 13,502 sika were bagged in County Wicklow. 
        However, there was 652 licences granted in County Wicklow, indicating that just under 21 sika were shot per deer hunter. 
        Naturally, the proportion of the total culled number varies between each hunter. 
        The majority of the 652 hunters shoot on average between 0 and 10 deer in the season, mainly for home consumption of venison.
        There are a few deer hunters associated with estates that would shoot over 80 deer in a season, though these would be in the minority. 
        It is difficult to even determine if the population of sika in the county is decreasing, increasing or is being maintained.
        We included the cull percentage range for sika deer in Wicklow to be in the \textit{Low} band, indicating that the herd population is expanding.
        As it is possible the population of sika deer in Wicklow is being maintained, we assess the impact of changing the mean cull percentage to that of the \textit{Mid} interval. 
        That is, we investigate the effect changing the mean cull percentage in County Wicklow from 10\% to 15\% for sika deer.
       
        The change in overall population estimates are shown in Table~\ref{tab:PopEstimatesMeanChange} when the mean cull percentage for sika deer in County Wicklow is changed.
        This increase in mean cull percentage may account for a proportion of the natural mortality numbers, it does not account for 
        the high levels of reproduction in yearling and adult female sika.
        \begin{table}[h]
            \caption{Comparison of wild deer population estimates and associated 95\% credible interval for mean cull rate and a change in one cull rate for sika deer (changed from 15\% to 10\%).}
            \label{tab:PopEstimatesMeanChange}
            \centering
            \begin{tabular}{@{}c c c c@{}}
                \toprule
                Cull percentages & Fallow & Red & Sika \\
                \midrule
                Mean & 150,384 [147,956 - 153,505] & 38,444 [37,339 - 39,555] & 180,706 [178,010 - 183,883] \\
                Altered & 150,441 [148,110 - 153,575] & 38,454 [37,414 - 39,534] & 135,683 [133,756 - 138,419] \\
                \botrule
            \end{tabular}
        \end{table}

        However, in addition to cull numbers and sightings, there are other forms of evidence that would suggest the population of sika deer in County Wicklow is increasing year on year. 
        County Wicklow is the highest forested county amongst all the 26 counties in the Republic of Ireland~\citep{ForestStats2022}, and the damage to commercial forests' timber yields would indicate a high density of sika deer~\citep{WoordlandsofIre, Murphy:2013ur}.
        There have also been reports on the absence of natural regeneration of flora and trees in (semi-)native woodlands and broad leaved forests, also indicative of a large sika population.
        Lastly, there have been cases on the adverse impacts and damage to agricultural crops and pasture whereby sika deer have fed on the spring grass before livestock is turned out after spending winter indoors. 
        
        The level of hunting, recreationally or professionally, is inadequate to control the population of sika deer in Wicklow. 
        The median population estimates from our model, at both 10\% and 15\% cull rates, still estimate an overabundance of sika in County Wicklow.
        For a favourable timber yield and to reduce the damage caused to a tolerable level, the ideal deer density is 2.5 deer per km$^2$~\citep{De-Nahlik:1992ux}. 
        County Wicklow covers an area of 2,027km$^2$, thus 5,068 deer for the entire county is the ideal number for tolerable levels of damage and for natural regeneration of flora to occur. 

    \subsection{Sensitivity to the Choice of Cull Percentages Sampled} \label{subsec:CullsSamp}
\begin{figure}[h]
            \centering
            \includegraphics[width=\textwidth]{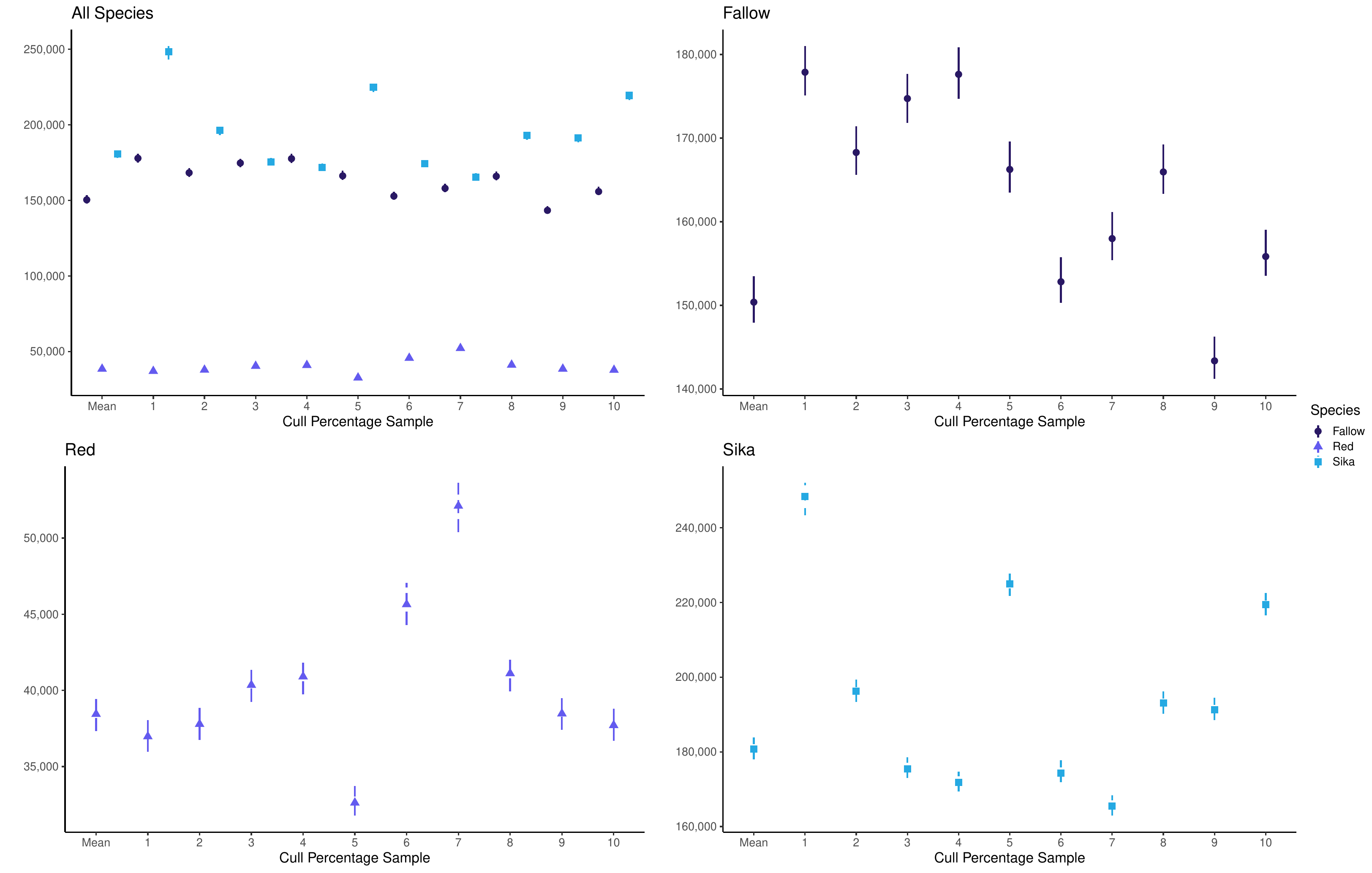}
                 \caption{11 sampled cull percentages and the associated 95\% credible intervals for estimated populations.}
                 \label{fig:Cull1Samp}
        \end{figure}

\begin{figure}[h]
            \centering
            \includegraphics[width = \textwidth]{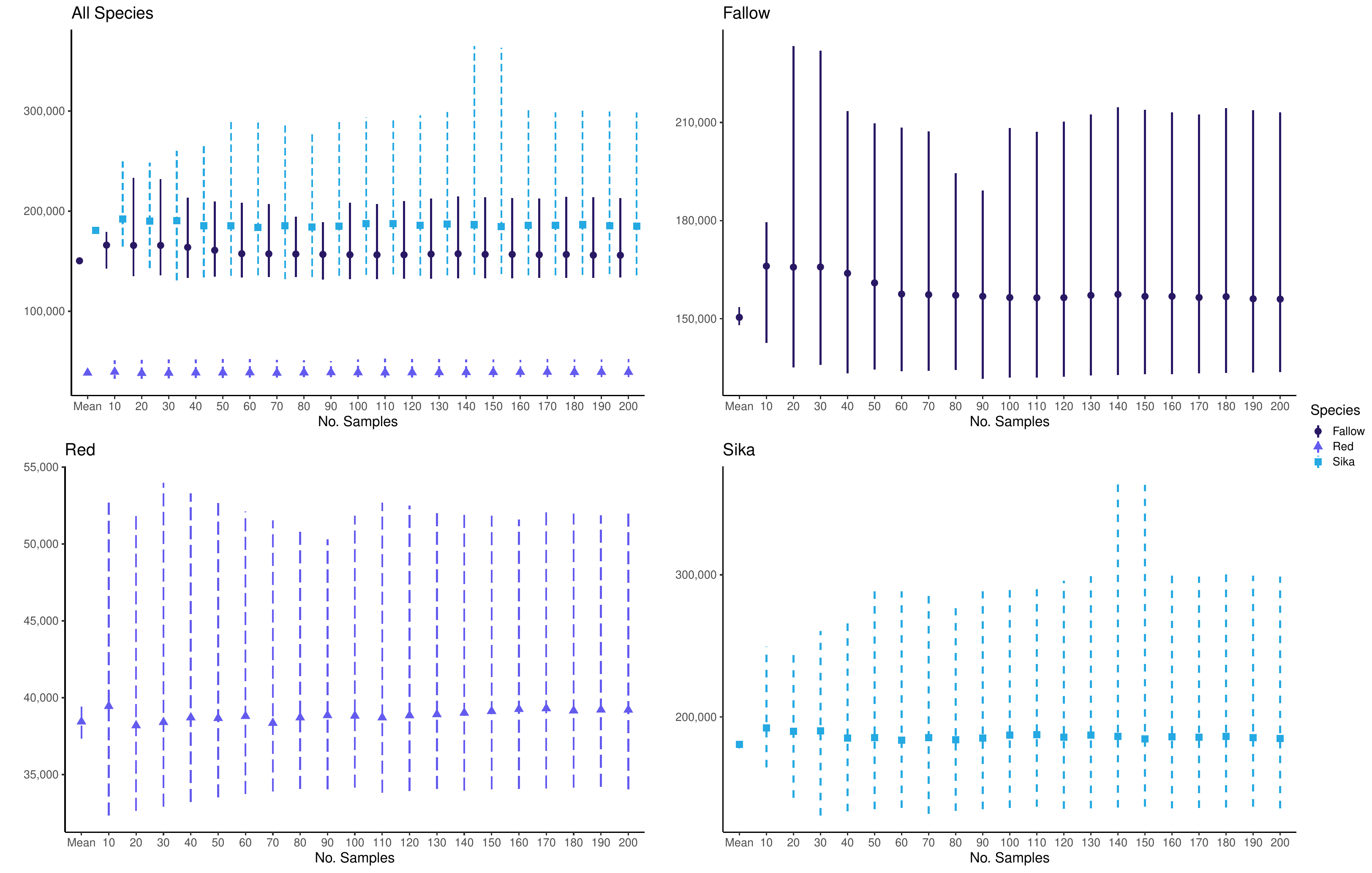}
            \caption{95\% credible intervals for various sample sizes of cull percentages.\\}
            \label{fig:rationalSampleCulls}
        \end{figure}

        To support our rational for sampling 200 cull percentages, we investigate the 95\% credible interval coverage of differing cull percentage sample sizes. 
        We initially investigate and compare the coverage provided by taking one cull percentage, shown in Figure~\ref{fig:Cull1Samp}.
        We then compare sampling various cull percentages to taking the mean cull percentage in Figure~\ref{fig:rationalSampleCulls}. 

        In Figure~\ref{fig:Cull1Samp} we highlight the different estimated medians and 95\% credible interval coverage provided by 11 different cull percentages, with one taking the mean cull percentage for each species, as used in Section~\ref{subsec:SampleVMean}. 
        Although the width of the interval is comparable between the 11 intervals within each species, the estimated median within each of these intervals differ. 
        Considering the inherent uncertainty we have in the cull percentages, it would be unwise to base our population estimates off one cull percentage.
        
        We therefore examine the differences in coverage when sampling 200 cull percentages from the intervals outlined in Table~\ref{tab:CullPrior}, to account for the additional variability in the cull percentages. 
        In Figure~\ref{fig:rationalSampleCulls} we show the 95\% credible interval coverage for the total population of each species for differing numbers of sampled cull percentages. 
        The top left plot shows the three species all on one plot, and the remaining three plots showcase the coverage obtained for each individual species. 
        Taking red deer as our example, shown in the bottom left plot, we can see a significant difference in the range covered between the mean cull percentage and sampling 10 cull percentages. 
        The increase in width provides further uncertainty in the population estimates, reflecting of our uncertainty in the cull percentages. 
        As the number of sampled cull percentages increase, the width of these intervals changes; however, there appears to be little change in the width and median estimate after 100 cull percentages sampled.

        Fallow deer present a different pattern to that seen in red deer. 
        With fallow deer, there is an increase in the median population estimate when comparing taking the mean cull percentages to sampling 10 cull percentages. 
        The median does not return to a similar value to that observed when taking the mean cull percentages. 
        That is the 95\% credible interval when taking the mean cull percentages does not contain the median population estimate when cull percentages are sampled. 
        This pattern is also observed for sika deer; however, it is less obvious.

    \subsection{Ecological Perspective and Population Estimate Validation} \label{subsec:EcologicalPerspective}
        Based on research from the UK, the wild deer population in the UK is estimated to be between 650,000 and 2 million~\citep{UKDeer}. 
        This is based on a max culled value of 350,000, representing a cull rate of between approximately 6\% and 15\% of the total population per annum. 
        Scaling this estimate to the Republic of Ireland, where approximately 42,000 deer were culled in the hunting season 2017/2018, we would expect the range of total deer populations to be between 78,000 and 240,000 based on these UK cull percentages. 
        However, our estimates from Section~\ref{sec:Results} (between 323,866 and 510,049 for the total population of all deer species) do not align with these values. 
        This could be for many reasons including the granularity of data sources and differing number of deer species between the two islands.
        Expert ecological opinion places cull numbers higher than those reported in official statistics, due to poaching and illegal hunting. 
        However, it is almost impossible to quantify the number of deer whose death can be attributed to illegal hunting, and therefore we do not include additional uncertainty for this.
        % Furthermore, expert ecological opinion is that the cull numbers in the Republic of Ireland are underestimated due to poaching and illegal hunting, and thus not reflected in official statistics. 

        We do however derive further confidence in our estimates from historical literature. 
        In 2000, O'Brien et al.~\cite{OBrien:2007tz} estimated the population of sika deer in Wicklow was 19,215.
        With a conservative 16\% increase in population numbers (sika increase their population numbers between 16-21\% annually~\citep{Kaji:2004vm}) year-on-year to 2018, this yields an estimated population of 277,116 sika for 2017/18.
        However, this estimate does not take into account the annual culling rate, which varies but increases from 2000 to 2018. 
        Our model may even be too conservative, but until such time that there are more accurate count numbers, we cannot elucidate accurate numbers, though we based our model assumptions on expert opinion and known data available. 
        % Our model may even be too conservative, but until such time that there are more accurate count numbers and associated data from County Wicklow for sika, we cannot elucidate accurate numbers, though we based our model assumptions on expert opinion and known data available. 
        
        The sika deer estimate does skew the estimates of the total deer population numbers in our model but the sika in Wicklow have a documented high fertility rate, 87\% of yearling females are pregnant and 85\% of adult females are pregnant and may reproduce one calf a year up to ages 12 years and older. 
        Very high survival rates of calves (85\%) also compound higher estimated numbers of the sika population in Wicklow~\citep{OBrien:2007tz}. 
        Similar high rates of fertility, pregnancy, and calf survival was also recorded in Killarney National Park in the past~\citep{ODonoghue:1991vx}.
        With an annual increase in population numbers between 16 and 21\%, it is possible that the numbers of sika are very high in 2017/18 and within the ranges of our model, based on initial population estimates of just over 19,000 sika in Wicklow in 2007.

\section{Discussion} \label{sec:Discussion}
    In this article we have introduced a novel N-mixture modelling framework for joint modelling of multiple species populations for ecological data presented at multiple differing spatial scales. 
    The sparse data provided from spatially replicated hunter visits on a 10km $\times$ 10km grid level are used to effectively estimate population sizes at that spatial resolution through the targeted downscaling of the aggregated deer cull data available at the county level, while also accounting for the detection and culling process. 
    Our model has the attractive aspect of allowing for estimation of inter-species abundance correlations, ensuring as much information as possible is extracted from the sparse data sets typically gathered by and available to ecological researchers. 
    All sources of uncertainty on model parameters are coherently modelled using a Bayesian framework, and incorporating best expert opinion from deer ecologists. 
    In terms of contributions to the ecological literature, this article presents the most complete proof of concept approach to the estimation of the populations of the three main species of wild deer in the Republic of Ireland given currently available data, and represents the first time their populations have been estimated with quantifiable uncertainty.  
    These population estimates are of interest to many stakeholders including conservationists, forest managers and policy makers, and our population estimates align with the expectations, or are deemed as plausible, by ecological experts within the Republic of Ireland. 
    This research also provides for a framework for assessing the impact of differing control strategies to limit deer numbers where overpopulation exists, as is suspected to be the case for sika deer numbers in the Wicklow county region.
    Due to the density of sika deer observed in County Wicklow, an increase in the culling percentage may be required to curtail damages and bring the population back to a manageable number. 
    
    Finally, we remark on some general research directions. It is straightforward to extend the presented model to account for additional sources of information. 
    Morera-Pujol et al.~\cite{Morera-Pujol:2023wj} use two data surveys conducted by the Irish forestry agency Coillte in their presence and relative abundance analysis. These included: (a) a survey completed by property managers on the likely presence of deer, and the species in the forests they maintain and (b) deer density surveys based on faecal pellet sampling in a subset of their properties between the years 2007 and 2020. 
    While the data collected by Coillte appears to be geo-tagged at a point referenced level, this can be aggregated to the spatial levels used within this study for seamless integration. This may assist in the refining of the deer hotspots from a location perspective, in addition to refining population estimates in areas where the number of deer spotted by hunters is low. Some citizen science data on a presence only basis has also been collected, however, while integrating citizen science data with other data sets can improve population estimates, some citizen science projects have poor data practices including a lack of accuracy, poor spatial or temporal representation, and insufficient sample size~\citep{Balazs:2021tm} and so it is not clear in an Irish context how these difficulties can be overcome.
    
    The data presented in this paper also typify those collected in many ecological studies and provide an avenue for the borrowing of information across multiple species data sets where inter-species correlation on abundances provide an approach to reduce uncertainties in population estimates. 
    It would be of interest to extend the approach to other non-ruminant animals, say to avian settings or similar. 
    Our model can also be extended to include temporal data, say through an autoregressive process, though this avenue was not applicable with the motivating data as the 10km $\times$ 10km observational data is only available for the 2017/2018 hunting season. 
    An additional future development of the model could be explore the impact of possible preferential sampling on species estimates~\citep{Diggle:2010vk}, as the observational data on deer is collected in areas where deer are likely to be observed. 
    The results of this research will be used to further develop strategies in future surveys, while also allowing informed discussions on wider species management and conservation programmes to manage the wild deer populations within the Republic of Ireland. 
    As mentioned in Section~\ref{subsec:DeerResults}, for 200 sampled cull percentages the model took approximately 11 days to run. 
    While this may test the bounds on feasibility for a purely spatial model, the addition of a temporal aspect may make the proposed model would make this model unusable. 
    It may be possible to speed-up computation with other methods such as recursive Bayes~\citep{Hooten:2021tj}. 

\backmatter

% \bmhead{Supplementary information}
%     If your article has accompanying supplementary file/s please state so here. 
%     Authors reporting data from electrophoretic gels and blots should supply the full unprocessed scans for key as part of their Supplementary information. This may be requested by the editorial team/s if it is missing.

%     Please refer to Journal-level guidance for any specific requirements.

\bmhead{Acknowledgements}
    This publication has emanated from research conducted with the financial support of Science Foundation Ireland under Grant number 18/CRT/6049. 
    For the purpose of Open Access, the author has applied a CC BY public copyright licence to any Author Accepted Manuscript version arising from this submission. 

\bmhead{Funding}
    This work was supported by Science Foundation Ireland Grant No. 18/CRT/6049 (A.~K.~H.).

\bmhead{Data and Code Availability}
    Data for this analysis was obtained from the Irish Deer Commission and National Parks and Wildlife Service. 
    The data was made available for research purposes and cannot be made publicly available. 
    Data sets used during the current study are available from R.~F.~C. on reasonable request.

\bmhead{Ethics approval and consent to participate}
    Not applicable. 
    
\bmhead{Author contribution}
    \begin{itemize}
        \item \textbf{Aoife K. Hurley}: Conceptualisation, development of statistical methodology, formal analysis, visualisation, data curation (processing), project administration, writing - original draft, writing - review \& editing.
        \item \textbf{Ruth F. Carden}: Conceptualisation, data curation (data collection, processing, and extraction), project administration, writing - review \& editing.
        \item \textbf{Sally Cook}: Data curation (processing and extraction), writing - review \& editing.
        \item \textbf{Irish Deer Commission}: Data curation (data collection, processing, and extraction).
        \item \textbf{Ferdia Marnell}: Data curation (data collection, processing, and extraction), writing - review \& editing.
        \item \textbf{Pieter A.J. Brama}: Writing - review \& editing.
        \item \textbf{Daniel J. Buckley}: Data curation (data collection, processing, and extraction), writing - review \& editing.
        \item \textbf{James Sweeney}: Conceptualisation, data curation (processing), writing - review \& editing.
    \end{itemize}
    
% \section*{Declarations}

% Some journals require declarations to be submitted in a standardised format. Please check the Instructions for Authors of the journal to which you are submitting to see if you need to complete this section. If yes, your manuscript must contain the following sections under the heading `Declarations':

% \begin{itemize}
% % \item Funding
% \item Conflict of interest/Competing interests (check journal-specific guidelines for which heading to use)
% \item Ethics approval and consent to participate
% \item Consent for publication
% % \item Data availability 
% \item Materials availability
% % \item Code availability 
% \end{itemize}

% \noindent
% If any of the sections are not relevant to your manuscript, please include the heading and write `Not applicable' for that section. 

% \bigskip

% Springer journals and proceedings: \url{https://www.springer.com/gp/editorial-policies}

% Nature Portfolio journals: \url{https://www.nature.com/nature-research/editorial-policies}

% \textit{Scientific Reports}: \url{https://www.nature.com/srep/journal-policies/editorial-policies}

% BMC journals: \url{https://www.biomedcentral.com/getpublished/editorial-policies}
% \end{flushleft}

\newpage
\begin{appendices}

\section{Additional RMSE Tables for the Estimated Spatial Surface}\label{sec:A1}
    This appendix includes the calculated RMSE for the spatial surface for each of the five simulated data sets for each of the eight retention levels. 
    The RMSE has been rounded to 3 decimal places for brevity. 
    \begin{table}[h]
        \caption{RMSE for eight retention levels for data set 1.}\label{tab:RMSE_C1}
        \centering
        \begin{tabular}{@{}l c c c@{}}
            \toprule
            Retention Level & Species 1 & Species 2 & Species 3 \\
            \midrule
            100\%   & 0.117 & 0.235 & 0.182 \\
            50\%    & 0.144 & 0.297 & 0.258 \\
            40\%    & 0.150 & 0.314 & 0.292 \\
            30\%    & 0.160 & 0.343 & 0.315 \\
            20\%    & 0.178 & 0.387 & 0.407 \\
            10\%    & 0.214 & 0.438 & 0.462 \\
            5\%     & 0.229 & 0.475 & 0.545 \\
            2.5\%   & 0.226 & 0.517 & 0.590 \\
            \botrule
        \end{tabular}
    \end{table}

    \begin{table}[h]
        \caption{RMSE for eight retention levels for data set 2.}\label{tab:RMSE_C2}
        \centering
        \begin{tabular}{@{}l c c c@{}}
            \toprule
            Retention Level & Species 1 & Species 2 & Species 3 \\
            \midrule
            100\%     & 0.122  & 0.202 & 0.160 \\
            50\%      & 0.160  & 0.255 & 0.220 \\
            40\%      & 0.174  & 0.280 & 0.233 \\
            30\%      & 0.188  & 0.295 & 0.281 \\
            20\%      & 0.224  & 0.323 & 0.311 \\
            10\%      & 0.256  & 0.369 & 0.381 \\
            5\%       & 0.293  & 0.402 & 0.430 \\
            2.5\%     & 0.340  & 0.434 & 0.494 \\
            \botrule
        \end{tabular}
    \end{table}

    \begin{table}[h]
        \caption{RMSE for eight retention levels for data set 3.}\label{tab:RMSE_C3}
        \centering
        \begin{tabular}{@{}l c c c@{}}
            \toprule
            Retention Level & Species 1 & Species 2 & Species 3 \\
            \midrule
            100\%     & 0.094 & 0.146 & 0.221 \\
            50\%      & 0.106 & 0.172 & 0.293 \\
            40\%      & 0.112 & 0.178 & 0.309 \\
            30\%      & 0.114 & 0.183 & 0.349 \\
            20\%      & 0.121 & 0.200 & 0.395 \\
            10\%      & 0.126 & 0.232 & 0.453 \\
            5\%       & 0.137 & 0.248 & 0.492 \\
            2.5\%     & 0.141 & 0.267 & 0.561 \\
            \botrule
        \end{tabular}
    \end{table}

    \begin{table}[h]
        \caption{RMSE for eight retention levels for data set 4.}\label{tab:RMSE_C4}
        \centering
        \begin{tabular}{@{}l c c c@{}}
            \toprule
            Retention Level & Species 1 & Species 2 & Species 3 \\
            \midrule
            100\%     & 0.082 & 0.164 & 0.306 \\
            50\%      & 0.093 & 0.194 & 0.350 \\
            40\%      & 0.099 & 0.205 & 0.388 \\
            30\%      & 0.104 & 0.202 & 0.424 \\
            20\%      & 0.111 & 0.236 & 0.454 \\
            10\%      & 0.128 & 0.257 & 0.533 \\
            5\%       & 0.132 & 0.277 & 0.611 \\
            2.5\%     & 0.149 & 0.300 & 0.602 \\
            \botrule
        \end{tabular}
    \end{table}

    \begin{table}[h]
        \caption{RMSE for eight retention levels for data set 5.}\label{tab:RMSE_C5}
        \centering
        \begin{tabular}{@{}l c c c@{}}
            \toprule
            Retention Level & Species 1 & Species 2 & Species 3 \\
            \midrule
            100\%     & 0.115 & 0.239 & 0.167 \\
            50\%      & 0.134 & 0.291 & 0.192 \\
            40\%      & 0.148 & 0.305 & 0.204 \\
            30\%      & 0.140 & 0.303 & 0.194 \\
            20\%      & 0.145 & 0.301 & 0.221 \\
            10\%      & 0.169 & 0.361 & 0.224 \\
            5\%       & 0.180 & 0.391 & 0.269 \\
            2.5\%     & 0.196 & 0.374 & 0.249 \\
            \botrule
        \end{tabular}
    \end{table}
    
% %%=============================================%%
% %% For submissions to Nature Portfolio Journals %%
% %% please use the heading ``Extended Data''.   %%
% %%=============================================%%

% %%=============================================================%%
% %% Sample for another appendix section			       %%
% %%=============================================================%%

% %% \section{Example of another appendix section}\label{secA2}%
% %% Appendices may be used for helpful, supporting or essential material that would otherwise 
% %% clutter, break up or be distracting to the text. Appendices can consist of sections, figures, 
% %% tables and equations etc.

\end{appendices}

\newpage
\bibliography{Deer_References}% common bib file

\end{document}